\documentclass[usenatbib]{mn2e}
\usepackage{multirow}
\usepackage{rotating}
\usepackage{colortbl}
\usepackage{color}
\usepackage[fleqn]{amsmath}
\usepackage{amssymb}
\usepackage{amsfonts}
\usepackage{verbatim}
\usepackage{scalefnt}
\usepackage[percent]{overpic}
\usepackage{siunitx}

\newcommand{\comments}[1]{} 

\title[Nuclear coups: dynamics of BHs in mergers]{Nuclear coups: dynamics of black holes in galaxy mergers}

\author[S. Van Wassenhove et al.]{Sandor Van Wassenhove,$^{1}$ Pedro R. Capelo,$^{1}$\thanks{E-mail: capelop@umich.edu} Marta Volonteri,$^{1,2}$\newauthor Massimo Dotti,$^{3}$ Jillian M. Bellovary,$^{4}$ Lucio Mayer$^{5}$ and Fabio Governato$^{6}$\\
$^1$Department of Astronomy, University of Michigan, Ann Arbor, MI 48109, USA\\
$^2$Institut d'Astrophysique de Paris, 98bis Boulevard Arago, F-75014 Paris, France\\
$^3$Dipartimento di Fisica G. Occhialini, Universit$\grave{a}$ degli Studi di Milano Bicocca, Piazza della Scienza 3, I-20126 Milano, Italy\\
$^4$Department of Physics and Astronomy, Vanderbilt University, Nashville, TN 37235, USA\\
$^5$Institute of Theoretical Physics, University of Z$\ddot{u}$rich, Winterthurerstrasse 190, CH-9057 Z$\ddot{u}$rich, Switzerland\\
$^6$Department of Astronomy, University of Washington, Box 351580, Seattle, WA 98195, USA}

\begin{document}

\maketitle

\begin{abstract}
We study the dynamical evolution of supermassive black holes (BHs) in merging galaxies on scales of hundreds of kpc to 10 pc, to identify the physical processes that aid or hinder the orbital decay of BHs. We present hydrodynamical simulations of galaxy mergers with a resolution of $\leq$20 pc, chosen to accurately track the motion of the nuclei and provide a realistic environment for the evolution of the BHs. We find that, during the late stages of the merger, tidal shocks inject energy in the nuclei, causing one or both nuclei to be disrupted and leaving their BH `naked', without any bound gas or stars. In many cases, the nucleus that is ultimately disrupted is that of the larger galaxy (`nuclear coup'), as star formation grows a denser nuclear cusp in the smaller galaxy. We supplement our simulations with an analytical estimate of the orbital-decay time required for the BHs to form a binary at unresolved scales, due to dynamical friction. We find that, when a nuclear coup occurs, the time-scale is much shorter than when the secondary's nucleus is disrupted, as the infalling BH is more massive, and it also finds itself in a denser stellar environment.

\end{abstract}



\section{Introduction}\label{ncp2013:sec:Introduction}
Observational evidence suggests that most massive galaxies contain black holes (BHs) with masses in the range of $10^6$--$10^9$~M$_{\odot}$ \citep{Ferrarese2005}. If many of these massive galaxies contain BHs, then galaxy mergers may lead to mergers between their central BHs. BH binaries are therefore expected to exist and one case has been confirmed, a 7-pc binary in the radio galaxy 0402+379 \citep{Rodriguez2006}, but candidate BH binaries remain rare and difficult to confirm. The coalescence of BHs provides a complementary mechanism for BH growth to accretion, which is also enhanced during galaxy mergers. In lower mass BHs (${M}_{\rm BH} \simeq 10^6$--$10^8$~M$_{\odot}$), accretion is the dominant mechanism for growth, but BH--BH mergers dominate in the highest mass BHs, which reside preferentially in gas-poor systems \citep{Malbon2007,2013arXiv1304.4583D,2013ApJ...768...29V}. BH mergers also lead to high signal-to-noise ratio bursts of gravitational waves, an important source for proposed space-based laser interferometers and the ongoing International Pulsar Timing Array \citep{Hobbs2010}, which will be sensitive to BH mergers.

For BHs to merge, however, they have to cross from distances of hundreds of kpc to sub-pc scale before one can be assured that they will merge within a Hubble time through emission of gravitational radiation. Several studies demonstrated that a gas-poor environment is unfavourable to rapid formation of BH binaries in galaxy mergers, and also to the shrinking of the orbit of two BHs below $\sim$pc scale for spherically symmetric systems (the `last parsec problem'; \citealt{BBR1980,milosavljevic2001}; for triaxial systems, see \citealt{Berczik2006,Khan2011,Preto2011,Gualandris_Merritt_2012,Khan2013,Vasiliev2013}). BH orbits in a gas-rich environment decay much faster, both on galactic and nuclear scales, due to efficient gravitational torques.

If the merging galaxies are not too dissimilar in mass (mass ratio $q \ge$ 1:10), the merger is likely to lead to the formation of a BH pair on $\simeq$100-pc scales \citep{VHM,stelios2005,Callegari2009}. We define a BH pair as two BHs residing in a single galaxy on scales of tens of pc to kpc. In a pair, the BHs are not bound to each other. When the BHs become bound to each other, they form a binary. This happens when the binary separation equals  $a_{\rm M}$, the radius at which the total enclosed mass is equal to twice the combined mass of the BHs: $M_{\rm tot}(r<a_{\rm M})=2M$. An alternative definition is that  the binary semimajor axis, $a_{\sigma}$, is the root of the equation $\sigma^2(r)-GM/r=0$, where $\sigma(r)$ is the velocity dispersion of the central remnant, $M$ is the combined mass of the BHs, and $G$ is the gravitational constant. These two definitions are equivalent if the galaxy mass distribution is described by a singular isothermal sphere. In all other cases, $a_{\sigma} < a_{\rm M}$, and we will show that this is the case for our mergers in Section \ref{ncp2013:sec:Discussion}. If a binary forms, it then continues to shrink under dynamical friction until the formation of a hard binary, when dynamical friction becomes inefficient \citep{Yu2002}. In gas-poor systems, the evolution of a hard binary is dominated by three-body interactions with nearby stars \citep{quinlan1996,2007ApJ...660..546S}. In gas-rich systems, friction against the gaseous background may continue to shrink the binary \citep{Escala2005,Dotti2007,Dotti2009,Mayeretal2006,Cuadra2009}. Once the binary reaches mpc scales, gravitational wave emission is efficient and the binary quickly coalesces. In the future, evidence for BH--BH binaries and mergers may instead come directly from detections of gravitational waves from the mergers themselves \citep[e.g.][]{polletto94,sesanaetal2004,PTA2008}. If the first step of this process is inefficient, however, then the subsequent steps do not occur, forming a bottleneck leading up to the formation of a BH binary. The merger of two galaxies does not ensure the merger of their BHs and it is vital to study the efficiency of the first step of the process: the formation of a BH pair. 

Numerous simulations have considered the triggering of BH accretion through equal-mass galaxy mergers \citep{DiMatteo2005,springel2005b,hopkins2006,Robertson2005,Johansson2009}. Several studies have also considered gas dynamics in minor mergers both with \citep{Younger2008} and without \citep{Cox2008} BHs. These studies have generally resolved scales of $\simeq$100 pc and focused on BH accretion and the evolution of galaxies along observed scaling relations, but not the dynamics of BH pairing and binary formation. Instead, it is assumed that BHs merge efficiently upon reaching the resolution limit of the simulation (Springel et al. 2005\nocite{springel2005b}). Additional mechanisms have been introduced in some studies to ensure efficient BH merging, including repositioning of BHs to the local potential minimum (Johansson et al. 2009\nocite{Johansson2009}) or the inclusion of a drag force acting on the BHs \citep{Younger2008}. \cite{Mayeretal2006} studied the formation of BH binaries in equal-mass mergers and found that in gas-rich merger remnants, BHs can sink and form a pc-scale binary on time-scales of Myr. \cite{stelios2005} and \cite{Callegari2009} instead focused on the dynamics of BH pairing in minor mergers. On smaller scales, the evolution of BH binaries in circumnuclear discs has been studied using idealized initial conditions \citep{Escala2005,Dotti2007,Dotti2009,Cuadra2009}. These simulations show that BH pairs can rapidly sink and form BH binaries in a gas-rich environment, but they sacrifice their link with the large-scale dynamics of the host galaxy in order to focus on the nuclear region with high (pc-scale) resolution.

Our simulations bridge the gap between large-scale, low-resolution simulations of galaxy mergers and the small-scale, high-resolution simulations of BH-binary evolution. By resolving $<$20-pc scales, we can accurately track the motion of the nuclei of the merging galaxies and study the efficiency of BH pairing in a realistic environment. Our simulations begin at $z = 3$, near the peak of the cosmic merger rate, when galaxy mergers were more common than at low redshift. We consider mergers meant to represent the most common mergers in the $\Lambda$ cold dark matter cosmology rather than relatively rare equal-mass mergers at $z = 0$ \citep[e.g.][]{Fakhouri2010}. We focus, therefore, on unequal-mass mergers with mass ratios of 1:2, 1:4, 1:6, and 1:10. We also study the effects of inclined and retrograde orbits.

In unequal-mass galaxy mergers the smaller galaxy, $G_2$, is prone to tidal stripping and tidal shocks from the larger one, $G_1$. These effects can completely disrupt $G_2$ early in a merger, stranding the secondary BH (BH$_2$) at kpc separations. However, strong star formation driven by nuclear torques in the secondary's nucleus ($N_2$) may lead to a reversal of this situation. If a dense stellar cusp forms around BH$_2$, tidal shocks may instead disrupt the primary's nucleus, $N_1$, causing a nuclear coup. This situation is more favourable to the formation of a BH pair compared to when $G_2$ or $N_2$ is disrupted.

We follow the interaction of the stellar nuclei on $<100$-pc scales and discuss the prospects for the formation of a BH binary. In Section \ref{ncp2013:sec:Numerical_setup}, we describe the numerical setup of our simulations. In Section \ref{ncp2013:sec:Dynamical_evolution}, we discuss in full detail the results of one of our runs, whereas in Section \ref{ncp2013:sec:Results} we generalize the analysis to the full suite of mergers. Finally, we compare our simulations and results to existing theoretical and observational work in Section \ref{ncp2013:sec:Discussion}.


\section{Numerical Setup}\label{ncp2013:sec:Numerical_setup}

In this section, we describe the numerical setup of our suite of merger simulations. It includes mergers of disc galaxies with mass ratios of 1:2, 1:4, 1:6, and 1:10, set at $z=3$, corresponding to the peak of the cosmic merger rate.


\subsection{Orbital parameters}\label{ncp2013:sec:Orbital_parameters}

We choose orbital parameters that match those of the most common halo mergers in cosmological simulations of galaxy formation \citep{Benson05}, where almost half of all mergers have an eccentricity $e$ between 0.9 and 1.1. \citet{Khochfar2006} find that 85 per cent of merging halo orbits have initial pericentre distances in excess of 10 per cent of the virial radius of $G_1$. Most simulations of galaxy mergers consider smaller pericentre distances instead, to save computational time, producing more direct collisions. We set instead the initial pericentre distance near 20 per cent of the virial radius of $G_1$, in order to be consistent with cosmological orbits. The initial separation between the galaxies is set near the sum of the two virial radii. We summarize the orbital parameters for each simulation in Table \ref{ncp2013:tab:params}.

We vary the angle between each galaxy's angular momentum axis and the overall orbital angular momentum vector, given by $\theta$ in Table \ref{ncp2013:tab:params}. We consider coplanar, prograde--prograde mergers, in which $\theta_1$ and $\theta_2$, the angles for $G_1$ and $G_2$, respectively, are both zero. In our inclined mergers, we set $\theta_1=\pi/4$ and $\theta_2=0$. Lastly, we consider coplanar, retrograde mergers, in which one of the galaxies is anti-aligned with the overall orbital angular momentum axis. In the coplanar, retrograde--prograde merger, $\theta_1 = \pi$ and $\theta_2 = 0$.  In the coplanar, prograde--retrograde merger, $\theta_1=0$ and $\theta_2 = \pi$.


\subsection{Galaxies}\label{ncp2013:sec:Galaxies}

\begin{table} \centering
\vspace{-2pt}
\caption[Simulation Parameters]{Orbital parameters for our simulations. $\theta_1$ and $\theta_2$ are the angles between the spin axis and the total orbital angular momentum axis for each galaxy. $q$ is the initial mass ratio between the merging galaxies. $e$ is the initial eccentricity of the orbit. $R_{\rm peri}$ is the first pericentre distance as a fraction of the virial radius of $G_1$. $R_{\rm init}$ is the initial separation divided by the sum of the virial radii of the merging galaxies.
\label{ncp2013:tab:params}}
\vspace{5pt}
{\small
\begin{tabular*}{0.4\textwidth}{cccccc}
\hline
Mass ratio ($q$) & $\theta_1$ & $\theta_2$ & $e$ & $R_{\rm peri}$ & $R_{\rm init}$ \\
\hline
1:2 & 0 & 0 & 1.02 & 0.3 & 1.05 \\
1:2 & $\pi/4$ & 0 & 1.02 & 0.225 & 1.05 \\
1:2 & $\pi$ & 0 & 1.02 & 0.225 & 1.05 \\
1:2 & 0 & $\pi$ & 1.02 & 0.225 & 1.05 \\
1:4 & 0 & 0 & 1.03 & 0.228 & 1.05 \\
1:4 & $\pi$/4 & 0 & 1.03 & 0.228 & 1.05 \\
1:6 & 0 & 0 & 1.03 & 0.228 & 1.05 \\
1:10 & 0 & 0 & 1.03 & 0.228 & 1.05 \\
\hline
\end{tabular*}
\vspace{10pt}
}
\end{table}

All the values in this section were chosen to be consistent with previous work \citep{Callegari2009,Callegari2011,VW2012}. Each galaxy is composed of a dark matter halo, a mixed stellar and gaseous disc, a stellar bulge, and a central massive BH (described in the next section). The dark matter halo is described by a spherical Navarro--Frenk--White profile \citep{NFW1996} with spin parameter $\lambda=0.04$. The dark matter halo concentration parameter is initialized to $c=3$. The disc has an exponential density profile with total mass equal to 4 per cent of the virial mass of the galaxy. The gas in the disc has a mass fraction $f_{\rm gas}=0.3$. Observations of high-redshift galaxies that are actively forming stars suggest that they may have higher gas fractions \citep{Tacconi2010}. The value used in this work represents more quiescent galaxies. The stellar bulge is described by a spherical \citet{Hernquist1990} density profile with total mass equal to 0.8 per cent of the virial mass of the galaxy. In each merger, $G_1$ has a virial mass of $2.24 \times 10^{11}$~M$_{\odot}$ (consistent with \citealt{adelberger2005b}), whereas the mass of $G_2$ scales according to the mass ratio.

For simplicity, each galaxy is initialized with solar metallicity and a uniform stellar population with an age of 2~Gyr to reflect the young age of the Universe at $z=3$. Without any existing feedback to heat the gas at the beginning of the simulation, much of the gas initially cools and vigorously forms stars. To avoid an unphysical burst of supernovae at the beginning of our merger simulations, we evolve the galaxies in isolation over 100~Myr (relaxation period), during which the star formation efficiency is gradually increased up to the value $c^*=0.015$.

In all the mergers of our suite, stellar particles have a mass of $3.3 \times 10^3$~M$_{\odot}$ and a softening length of 10~pc, whereas gaseous particles have a mass of $4.6 \times 10^3$~M$_{\odot}$ and a softening length of 20~pc. The dark matter particle mass was instead chosen as a function of the BH mass, the maximum dark matter particle mass being set to 1/7 of the smallest BH mass in the merger, to limit excursions of BHs from the centre of each galaxy. For the 1:2 and 1:4 simulations, the mass and softening length were set to $1.01 \times 10^5$~M$_{\odot}$ and 30~pc, respectively. For the 1:6 and 1:10 simulations, the dark matter particle masses and softening lengths were lowered to reflect the low mass of the BH in the secondary galaxy. The 1:6 simulation used a dark matter particle mass of $7.56 \times 10^4$~M$_{\odot}$ and softening length of 27~pc. The 1:10 simulation used a dark matter particle mass of $3.9 \times 10^4$~M$_{\odot}$ and softening length of 24~pc.

We performed all our simulations using the $N$-body smoothed particle hydrodynamics code {\scshape gasoline} \citep{wadsley04}, an extension of the pure gravity tree code {\scshape pkdgrav} \citep{stadel01}. {\scshape gasoline} includes explicit line cooling for atomic hydrogen, helium and metals, as well as a physically motivated prescription for star formation, supernova feedback and stellar winds \citep{Stinson2006}. In particular, stars are allowed to form if the parent gas particle is colder than 6000~K and denser than 100~a.m.u.~cm$^{-3}$, and supernovae release $10^{51}$~erg into the surrounding gas, according to the blast wave formalism of \citet{Stinson2006}.


\subsection{Black holes}\label{ncp2013:sec:Black_holes}

A recent implementation in the {\scshape gasoline} code has been the inclusion of a recipe for BH physics \citep{Bellovary10}, in which BHs are implemented as sink particles that accrete from nearby gas particles according to an Eddington-limited Bondi--Hoyle--Littleton accretion formula. BH accretion gives rise to feedback, implemented as thermal energy injected into the nearest gas particle according to $\dot{E} = \epsilon_{\rm f} \epsilon_{\rm r} \dot{M}_{\rm BH} c^2$, where $c$ is the speed of light in vacuum, $\epsilon_{\rm r}=0.1$ is the radiative efficiency and $\epsilon_{\rm f}$ is the feedback efficiency, chosen to be equal to 0.001 to match the local M$_{\rm BH}$--M$_{\rm bulge}$ relation at the end of the merger.

We place a single BH at the centre of each galaxy, after the galaxy has been initialized. Its mass is set according to the local M$_{\rm BH}$--M$_{\rm bulge}$ relation \citep{MarconiHunt2003}. The mass of the primary BH (BH$_1$) in each simulation is initially set to $3 \times 10^6$~M$_{\odot}$, whereas BH$_2$ has a mass proportional to the mass ratio between the galaxies, producing a minimum initial mass of $3 \times 10^5$~M$_{\odot}$ in the 1:10 merger. The softening length of all BHs is set to 5~pc, regardless of their mass.


\section{Dynamical evolution}\label{ncp2013:sec:Dynamical_evolution}

In this section, we describe physical processes influencing the dynamics of galaxy mergers. We highlight the processes that modify the gaseous and stellar content of galaxies. The removal or addition of gas and stars affects the overall orbital decay and, in particular, the evolution of the nuclei and their embedded BHs. Here the nucleus of each galaxy refers to the material within $\simeq$100~pc of the centre of the galaxy. The presence or absence of a dense nucleus surrounding the BH is crucial to the dynamics and eventual formation of a BH binary \citep{Yu2002}.

We find that ram pressure (Section~\ref{ncp2013:sec:Ram_pressure}) and tidal stripping (Section~\ref{ncp2013:sec:Tidal_stripping}) are important only on large scales, stripping $G_2$ of its gas and hindering its ability to retain gas for nuclear star formation. On the other hand, tidal torques (Section~\ref{ncp2013:sec:Star_formation_driven_by_tidal_torques}) are important in driving nuclear star formation in $G_2$, helping create a dense nucleus. Owing to our high spatial resolution, we are able to isolate the crucial importance of tidal heating (Section~\ref{ncp2013:sec:Tidal_heating}) at late times, as the energy exchanged from the nuclei during close pericentre passages is what eventually determines the disruption of one, the other, or possibly both, nuclei.

We use the 1:4 coplanar, prograde--prograde merger to illustrate the general properties and phases of the merger. In Section \ref{ncp2013:sec:Results}, we discuss the remaining simulations and how they differ from the general picture presented here.

\begin{figure}
\centering
\vspace{5pt}
\includegraphics[width=1.0\columnwidth,angle=0]{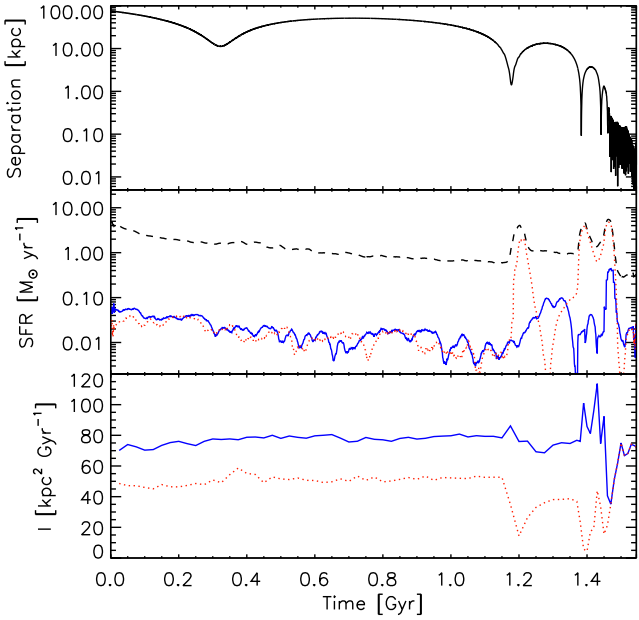}
\vspace{6pt}
\caption[Results of the 1:4 coplanar, prograde--prograde simulation]{Results of the 1:4 coplanar, prograde--prograde merger. Top panel: separation between the central BHs of each galaxy. Middle panel: global SFR across both galaxies (black, dashed line), and central ($<$100~pc) SFR of $G_1$ (blue, solid line) and $G_2$ (red, dotted line). Bottom panel: angular momentum per unit mass of gas in the central kpc of $G_1$ (blue, solid line) and $G_2$ (red, dotted line). All quantities are shown as a function of time.}
\label{ncp2013:fig:3panels_1to4_cop_pro_pro}
\end{figure}


\subsection{Ram pressure}\label{ncp2013:sec:Ram_pressure}

When the gaseous discs of the galaxies collide, they do not pass through each other as the stars and dark matter do, but feel pressure from the gas in the opposing disc. The collisions dissipate the orbital energy of the gas in the galaxies, creating the gaseous bridge that links the galaxies after the second pericentre passage. We consider the effects of ram pressure from $G_1$'s disc on $G_2$'s disc \citep{GunnGott,MovdBW}:

\begin{equation}\label{ncp2013:eq:P_ram1}
P_{\rm ram}=\rho_{1}v^2 > 2\pi G \Sigma_{*,2}(R)\Sigma_{\rm ISM,2}(R),
\end{equation}

\noindent where $\rho_{1}$ is the gas density of $G_1$'s disc, $v$ is the relative velocity between the galaxies during the collision, and $\Sigma_{*,2}(R)$ and $\Sigma_{\rm ISM,2}(R)$ are the stellar and gaseous surface densities in $G_2$ at a radius $R$. If the inequality in equation \eqref{ncp2013:eq:P_ram1} is satisfied at a given radius $R$, then the gas in $G_2$'s disc at that radius will be stripped during the collision.

This prescription for ram pressure is generally used to describe ram pressure from a hot, low density medium, whereas we are considering direct collisions between cold, dense gas clouds. The gaseous discs are inhomogeneous and the overall collision is short, lasting $\simeq$50~Myr. None the less, equation \eqref{ncp2013:eq:P_ram1} is instructive. To illustrate how the impact of ram pressure varies with the mass ratio of the merging galaxies, we rewrite equation \eqref{ncp2013:eq:P_ram1} using the surface densities in our galaxy models \citep{MoMaoWhite1998}:

\begin{equation}\label{ncp2013:eq:P_ram2}
P_{\rm ram} > \frac{Gf_{\rm g,2}(1-f_{\rm g,2})M_{\rm d,2}^2}{2\pi R_{\rm d,2}^4}{\rm e}^{-2R/R_{\rm d,2}} \propto M_{\rm d,2}^{2/3}{\rm e}^{-2R/R_{\rm d,2}}.
\end{equation}

Here $f_{\rm g,2}$ is the gas fraction of $G_2$, and $M_{\rm d,2}$ and $R_{\rm d,2}$ are the mass and scale radius of $G_2$'s disc, respectively. As the mass ratio of the merger decreases, $M_{\rm d,2}$ is lower and a given $P_{\rm ram}$ strips $G_2$ down to a smaller radius. This is primarily because the stellar and gaseous surface densities of $G_2$ decrease as the mass of the galaxy decreases, leaving it less resistant to ram pressure.

\begin{figure}
\centering
\vspace{5pt}
\includegraphics[width=0.90\columnwidth,angle=0]{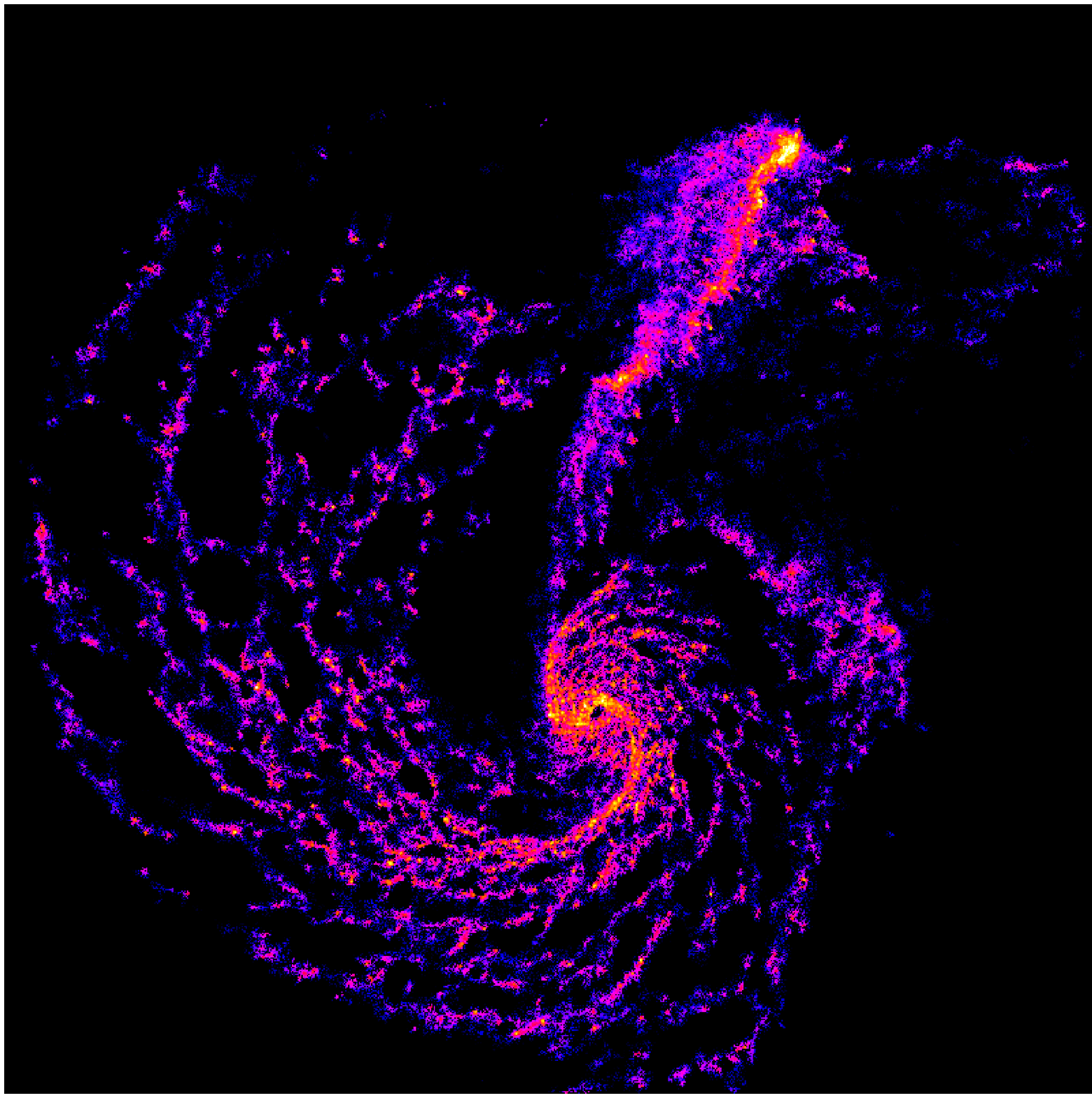}
\vskip 0.5mm
\includegraphics[width=0.90\columnwidth,angle=0]{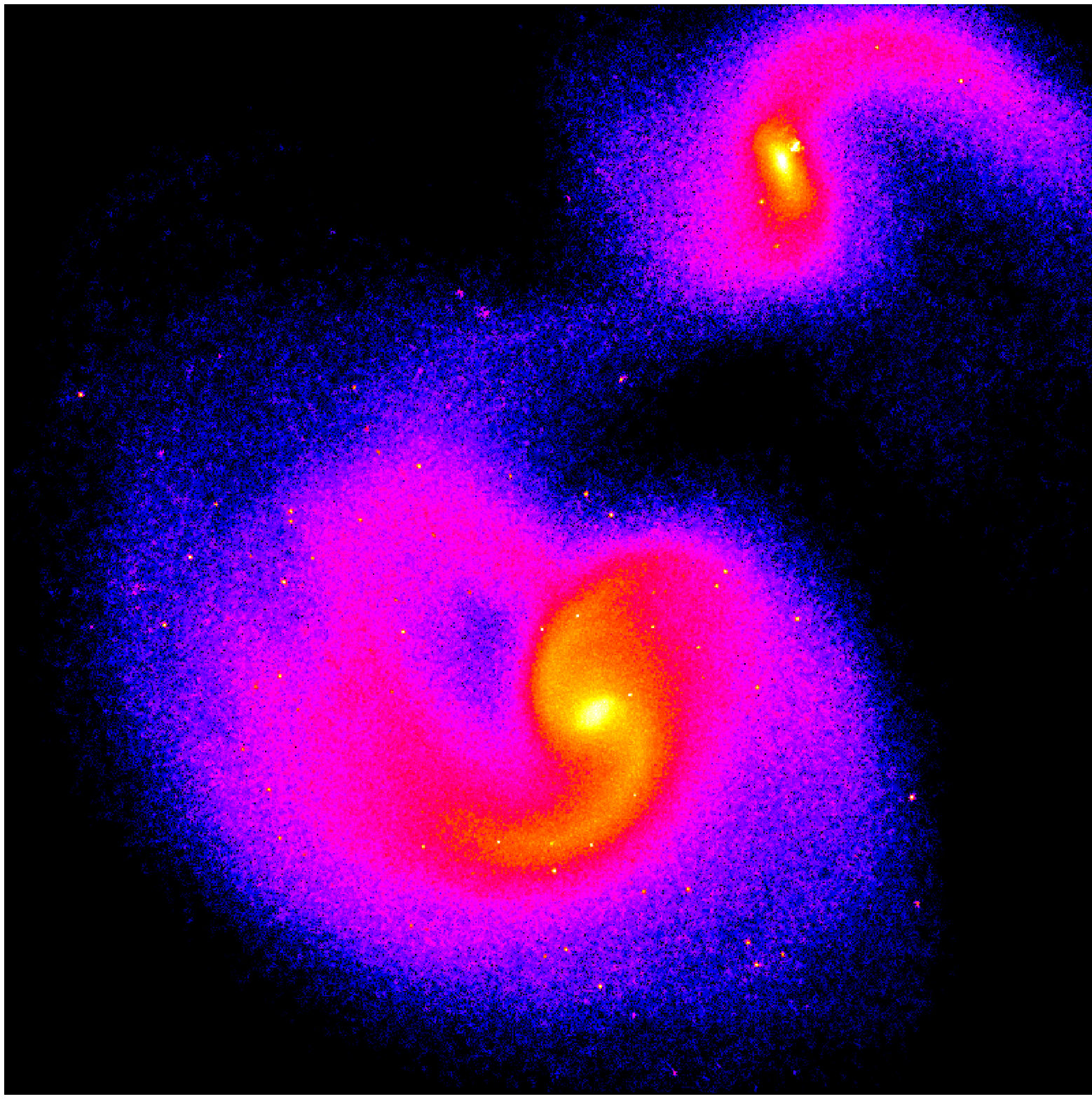}
\vskip 1.1mm
\begin{overpic}[width=0.89\columnwidth,angle=0]{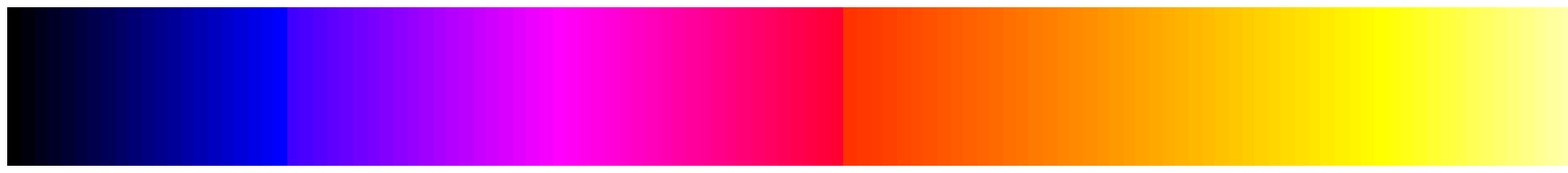}
\put (1,3) {\textcolor{white}{$10^2$}}
\put (48,3) {$10^4$}
\put (95,3) {$10^6$}
\end{overpic}
\vspace{20pt}
\caption[Gas and stellar density snapshots]{Gas (upper panel) and stellar (lower panel) density snapshot in the 1:4 coplanar, prograde--prograde merger at $t =1.2$~Gyr, just following the second pericentre passage. $G_1$ is at the bottom and $G_2$ is at the top. The distance between the centres of the two galaxies is 6.3~kpc. The colour bar shows the (logarithmic) density scale in units of $2.2 \times 10^5$~M$_{\odot}$~kpc$^{-3}$.}
\label{ncp2013:fig:gas_and_stellar_map_1to4_cop_pro_pro}
\end{figure}

In the 1:4 coplanar, prograde--prograde merger, $G_1$'s disc is relatively unaffected by ram pressure, whereas the outskirts of $G_2$ are strongly stripped (see the upper panel of Fig.~\ref{ncp2013:fig:gas_and_stellar_map_1to4_cop_pro_pro} for a map of the gas density following second pericentre). Gaseous inflows increase the central surface density of $G_2$ by a factor of 5 or more, helping the central gas to survive the interaction with $G_1$. Immediately following second pericentre, $\simeq$45 per cent of the gas in the central 100~pc of $G_2$ originated in the disc of $G_1$, suggesting that $G_2$ efficiently captures gas during the collision. While the low density gas in the outskirts of $G_2$ is stripped, forming a bridge between the galaxies, the dense central gas survives the encounter. $G_2$ captures gas as it plows through $G_1$'s disc, similarly to what discussed in Callegari et al. (2009), but well before circularization of the orbit. We also see evidence of compression in the central gas of $G_2$ due to ram pressure during and immediately following the second pericentre passage. The pressure of the nuclear gas [$P = k_{\rm B} \rho T/(\mu m_{\rm u})$, where $k_B$ is the Boltzmann constant, $\mu$ and $T$ are the mean molecular weight and temperature of the gas, respectively, and $m_u$ is the atomic mass unit] increases by three orders of magnitude, reaching a value corresponding to $P_{\rm ram}$ from cold, dense gas in $G_1$'s disc ($\rho \simeq 10^3$--$10^4$~a.m.u.~cm$^{-3}$; $v = 500$~km~s$^{-1}$ at second pericentre). Numerous simulations of ram pressure from a hot, low density medium have suggested that it can enhance star formation in the disc and wake of the stripped galaxy \citep{Evrard91,Vollmer2001,Kronberger2008,Kapferer2009}.

The effects of ram pressure will be maximized for our coplanar mergers, where both gaseous discs must pass completely through each other. The rotation of the galaxies can also increase the impact of ram pressure if the galaxies rotate into the collisional interface, increasing the velocity $v$ in $P_{\rm ram} = \rho v^2$ (see our coplanar, retrograde mergers in Section \ref{ncp2013:sec:Impact_of_orbital_parameters}).

\subsection{Star formation driven by tidal torques}\label{ncp2013:sec:Star_formation_driven_by_tidal_torques}

During close pericentre passages between the galaxies, gravitational torques between the galaxies lead to the formation of stellar and gaseous bars. The gaseous bar tends to lead the stellar bar, causing a torque upon the gas that removes angular momentum \citep{MihosHernquist1996}. The angular momentum loss in the gas causes gaseous inflows from kpc scales into the nuclear region. The bottom panel of Fig.~\ref{ncp2013:fig:3panels_1to4_cop_pro_pro} shows the angular momentum per unit mass in the central kpc of each galaxy in the 1:4 coplanar, prograde--prograde merger. We focus on the angular momentum in the central kpc rather than in the central 100~pc because large-scale inflows are important for funnelling gas into the central regions of each galaxy. In agreement with the findings of \cite{MihosHernquist1996}, we find that the presence of a bulge stabilizes each galaxy against the formation of a bar instability during the first pericentre passage. Accordingly, there is no loss of angular momentum in the gas. At second pericentre and at subsequent pericentre passages, however, torques lead to strong angular momentum loss and gaseous inflows. The response of $G_1$ is considerably weaker than that of $G_2$, with $G_1$'s disc losing little angular momentum until late in the merger. The relatively more massive $G_1$ produces a strong tidal field and it is not significantly perturbed by $G_2$'s weaker tidal field.

Inflowing gas fuels star formation in each galaxy. The strongest gaseous inflows and corresponding bursts of star formation occur during pericentre passages, when tidal torques between the galaxies are strongest. At first pericentre, however, the presence of a bulge stabilizes the galaxies and there are neither inflows nor any enhancement in star formation (middle panel of Fig.~\ref{ncp2013:fig:3panels_1to4_cop_pro_pro}). Instead, the galaxies evolve quiescently until the second pericentre passage at $t \simeq 1.2$~Gyr. The global star formation rate (SFR) decreases initially as the galaxies continue to settle from the initial conditions. Once the galaxies have settled, the SFR gradually falls as gas is depleted through star formation. During this initial, quiescent phase of the merger, the nuclear SFR in each galaxy is low, remaining at approximately two orders of magnitude less than the global SFR.

At second pericentre passage, tidal torques remove angular momentum from the gas in $G_2$, driving inflows and building up a high central gas density. Unlike at first pericentre, the gas discs collide and the gas is shocked and dissipates its orbital energy. Fig.~\ref{ncp2013:fig:gas_and_stellar_map_1to4_cop_pro_pro} shows a snapshot of the gas and stellar densities just after second pericentre. The collision causes much of the gas in $G_2$ to lag behind the stellar component in the form of a gaseous bridge. This bridge contains significant cold gas and hosts moderate star formation, in agreement with observations of molecular gas in bridges resulting from disc collisions \citep{Braine2004,Lisenfeld2008,Vollmer2012}. The high density gas in the centre of $G_2$ survives the encounter and is compressed due to ram pressure during the collision, forming a small (radius $\simeq$100~pc) clump of star-forming gas.

The dense central clump of gas in $G_2$ hosts a burst of star formation following second pericentre, reaching a rate of 4.4~M$_{\odot}$~yr$^{-1}$ which is a five hundred fold increase over the quiescent central SFR of $\simeq$0.01~M$_{\odot}$~yr$^{-1}$. At its peak, the central 100-pc region of $G_2$ is hosting $\simeq$80 per cent of the global star formation compared to 1 per cent of the global rate previously, showing how effectively the close encounter has concentrated the gas there. The starburst lasts $\simeq$25~Myr before supernova feedback halts any further star formation. $G_1$, on the other hand, experiences weak inflows immediately following second pericentre and shows no significant increase in star formation.

As the galaxies separate and approach second apocentre passage, $G_1$ develops a weak bar instability. The bar funnels gas into the centre of the galaxy, but the overall loss of angular momentum is small and the nuclear star formation is far weaker than that of $G_2$ at second pericentre passage. Meanwhile, $G_2$ reforms a small (radius $\simeq$800~pc) gaseous disc from gas in the bridge and tidal features, including a significant amount of gas that originally resided in $G_1$. The new disc forms with the opposite angular momentum of the previous one, turning the third pericentre passage into a prograde--retrograde encounter.

At third pericentre, angular momentum loss drives further gaseous inflows in $G_2$. The central regions are again compressed during the collision with the more massive and extended gaseous disc of $G_1$. This compression increases the density of the central gas, driving another burst of star formation in $G_2$. The nuclear SFR in the central 100~pc reaches 7.7~M$_{\odot}$~yr$^{-1}$, with 92 per cent of the global star formation occurring there during the burst. As at second pericentre, the response of $G_1$ is far weaker and there is no significant gaseous inflow or star formation.

During the remainder of the merger, $G_2$ does not leave the disc of $G_1$. The remaining pericentre passages occur much more quickly than the early passages, leaving little time for $G_2$ to reform a dense gaseous disc. The central SFR in $G_2$ remains high at $>$0.5~M$_{\odot}$~yr$^{-1}$, but there are no strong bursts at the fourth and fifth pericentre passages. The last peak of star formation occurs in the merger remnant following the sixth pericentre passage as the remaining gas in both galaxies engages in a starburst. This last starburst yields the highest SFRs of the entire simulation, with the global rate reaching 10.5~M$_{\odot}$~yr$^{-1}$, but it occurs after the stellar nuclei have merged and does not contribute to the formation of a pre-merger central cusp.

\begin{figure}
\centering
\vspace{5pt}
\includegraphics[width=1.0\columnwidth,angle=0]{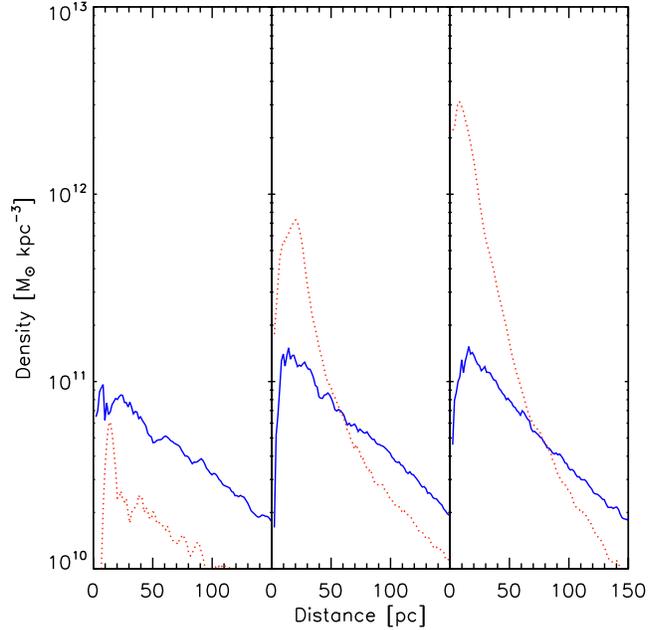}
\vspace{6pt}
\caption[Total density profile in the 1:4 coplanar, prograde--prograde merger]{Total density of stars, gas, and dark matter, as a function of distance $r$ from the central BH of each galaxy, for $G_1$ (blue, solid line) and $G_2$ (red, dotted line) in the 1:4 coplanar, prograde--prograde merger. Left-hand panel: $t = 1$~Gyr, before second pericentre. Middle panel: $t = 1.3$~Gyr, after second pericentre. Right-hand panel: $t = 1.42$~Gyr, after third pericentre. At each time, $r = 0$ corresponds to the position of the central BH of the given galaxy.}
\label{ncp2013:fig:central_density_1to4_cop_pro_pro}
\end{figure}

Fig.~\ref{ncp2013:fig:central_density_1to4_cop_pro_pro} shows the total density of stars, gas, and dark matter in each galaxy at three different times, as a function of distance from the central BH of each galaxy. The left-hand panel shows $t = 1$~Gyr, prior to the second pericentre, when neither galaxy has experienced any strong merger-driven star formation. At this time, $G_2$ is less dense than $G_1$, as was the case in the initial conditions. The middle panel shows $t = 1.3$~Gyr, near apocentre following the second pericentre. Both galaxies have built up a denser central cusp through new star formation, but the nuclear starburst in $G_2$ at second pericentre has left $N_2$ significantly denser. The right-hand panel shows the density profiles at $t = 1.42$~Gyr, after third pericentre, when the majority of the central star formation in both galaxies is complete. After continued strong star formation following the third pericentre, $G_2$ remains denser on small scales, $r \leq 75$~pc.

Not all star formation that contributes to the build-up of the nuclear cusp is local. Even during pericentre passages, there is a significant amount of star formation outside the nuclei. The off-centre gas participating in the starbursts tends to be dense and clumped, yielding clusters of new stars. Some of these clusters will sink to the centre of the nuclei under the effects of dynamical friction and contribute to the nuclear stellar population.

Efficient nuclear star formation in $G_2$ yields a stellar cusp that is denser than that of $G_1$. The additional mass in new stars ensures the survival of $N_2$, aiding in the formation of a BH pair. To understand the continued evolution of the predominantly stellar nuclei as they merge, we consider the effects of tidal stripping and tidal heating and determine whether they can account for the behaviour seen in our simulations.

\subsection{Tidal stripping}\label{ncp2013:sec:Tidal_stripping}

In a slow encounter between the two galaxies, the static tidal field produced can remove material from each galaxy outside a limiting tidal radius. Observationally, the effects of tidal stripping are commonly seen in globular clusters and dwarf galaxies \citep[e.g.][]{King1962}.

The natural time-scale for tidal stripping is the orbital time-scale of the stars in the satellite at its tidal radius. The tidal fields of the galaxies are important on large scales for mass loss, particularly for the gaseous bridge that links the discs following second pericentre. $G_2$ can only reform its disc from gas that remains bound following the disc collision. On small scales, the stellar nuclei are unaffected by tidal stripping. The pericentre passages last an order of magnitude less than the relevant orbital time-scales, suggesting that there is insufficient time for tidal stripping to act on the nuclei. During the late stages of the merger, we instead consider the impact of fast encounters through tidal shocks.


\subsection{Tidal heating}\label{ncp2013:sec:Tidal_heating}

\begin{figure}
\centering
\vspace{5pt}
\includegraphics[width=1.0\columnwidth,angle=0]{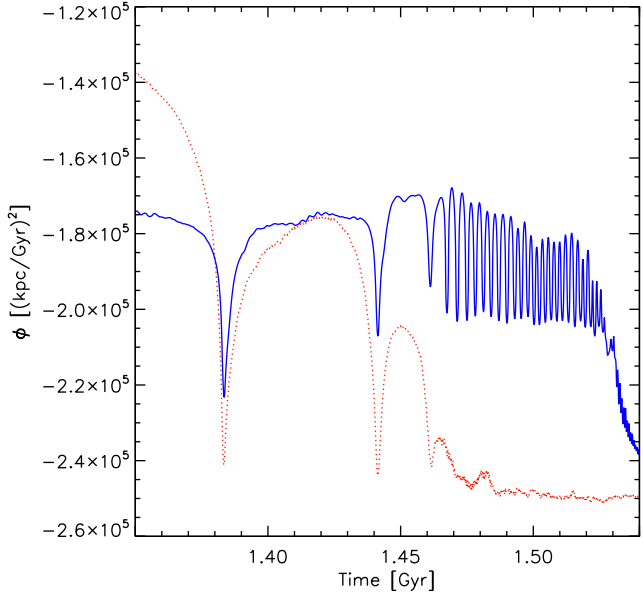}
\vspace{6pt}
\caption[]{Gravitational potential of the BHs originally in $G_1$ (blue, solid line) and in $G_2$ (red, dotted line) at late times, in the 1:4 coplanar, prograde--prograde merger. The signature of a nuclear coup is visible at $t \simeq 1.45$~Gyr, when BH$_2$ becomes the most tightly bound object. Note how BH$_1$ later progresses to meet the new central BH, BH$_2$.}
\label{ncp2013:fig:BH_potential_vs_time}
\end{figure}

\begin{figure}
\centering
\vspace{6pt}
\includegraphics[width=0.45\columnwidth,angle=0]{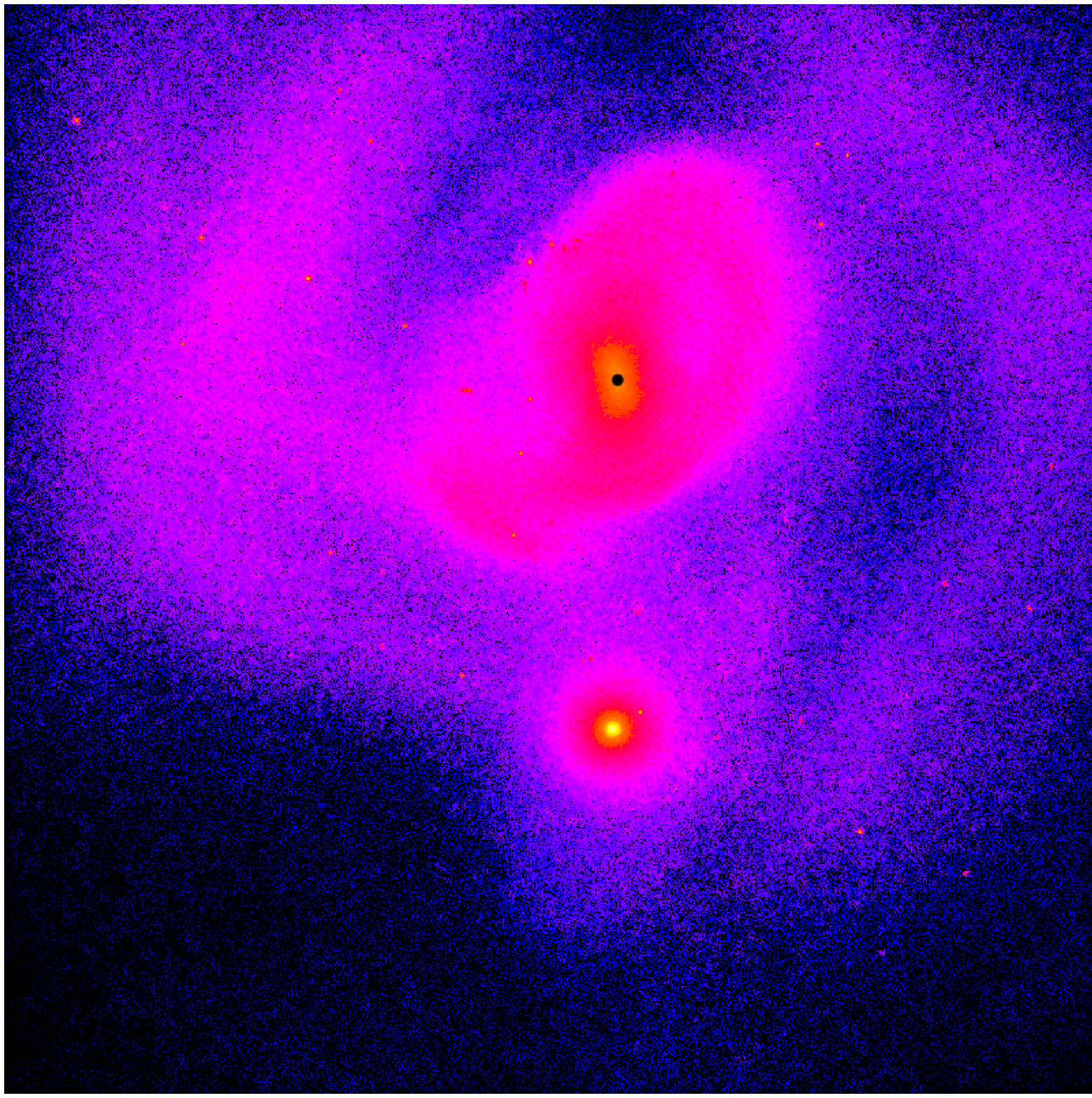}
\includegraphics[width=0.45\columnwidth,angle=0]{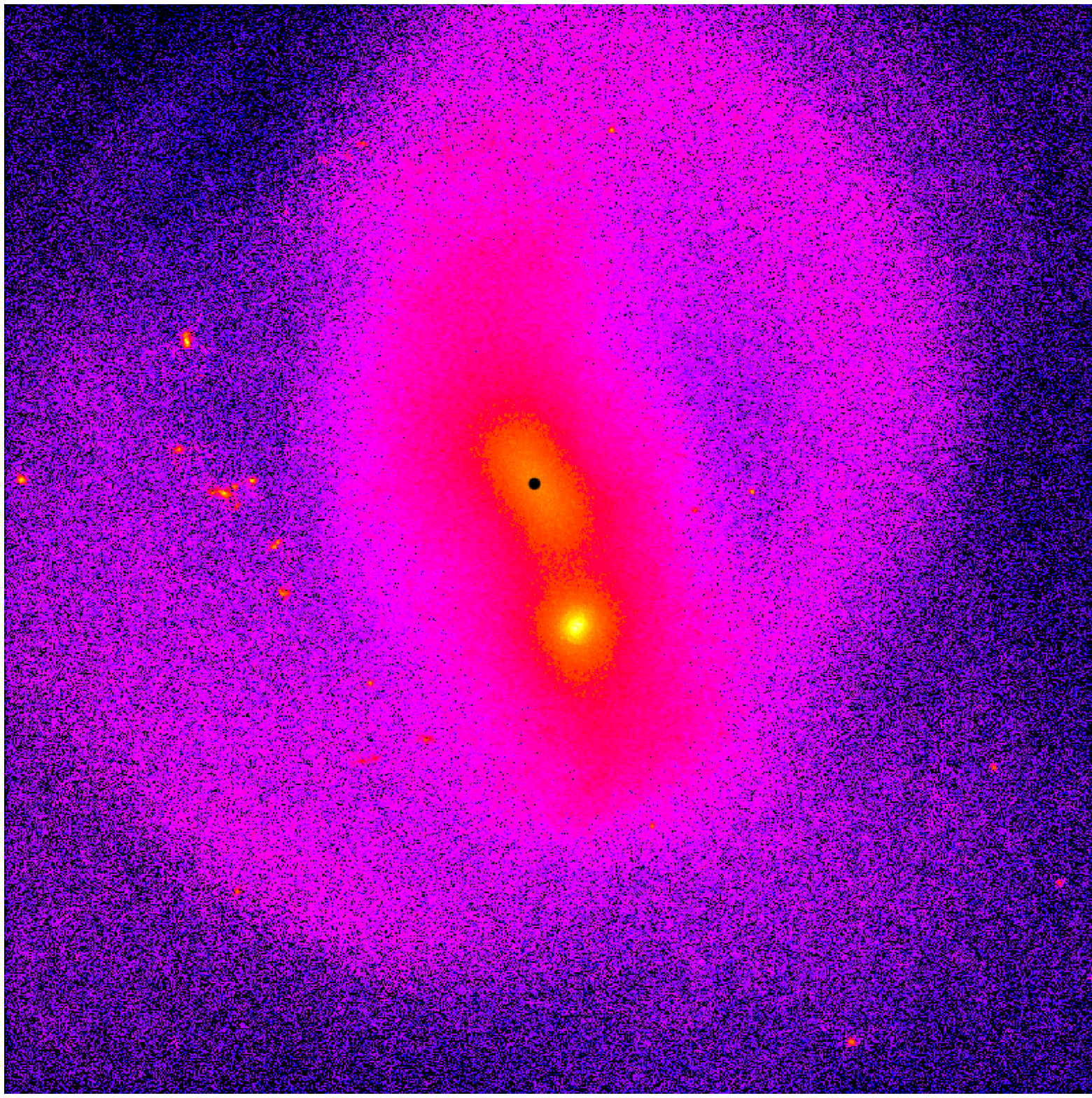}
\vskip 0.5mm
\includegraphics[width=0.45\columnwidth,angle=0]{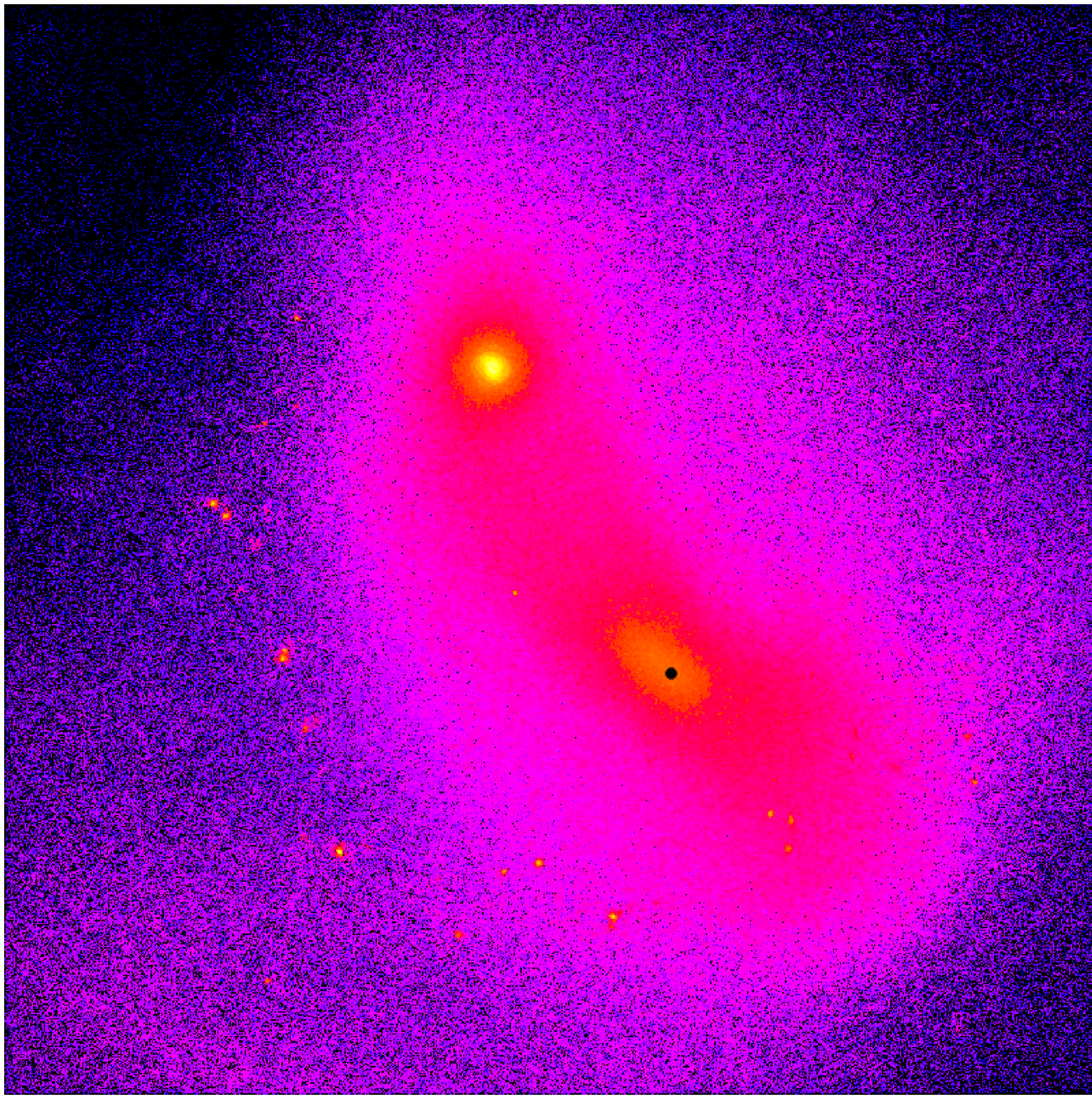}
\includegraphics[width=0.45\columnwidth,angle=0]{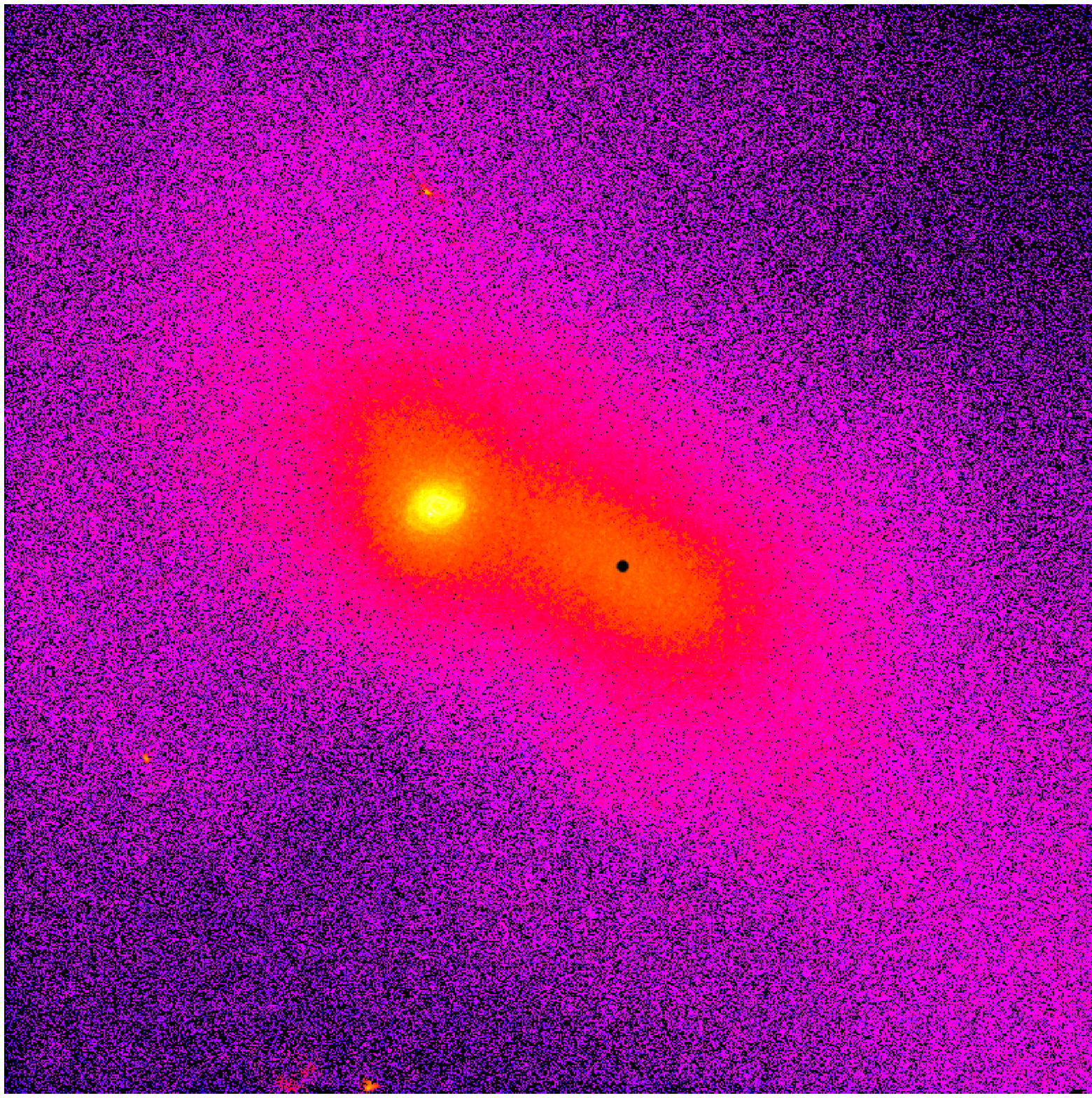}
\vskip 0.5mm
\includegraphics[width=0.45\columnwidth,angle=0]{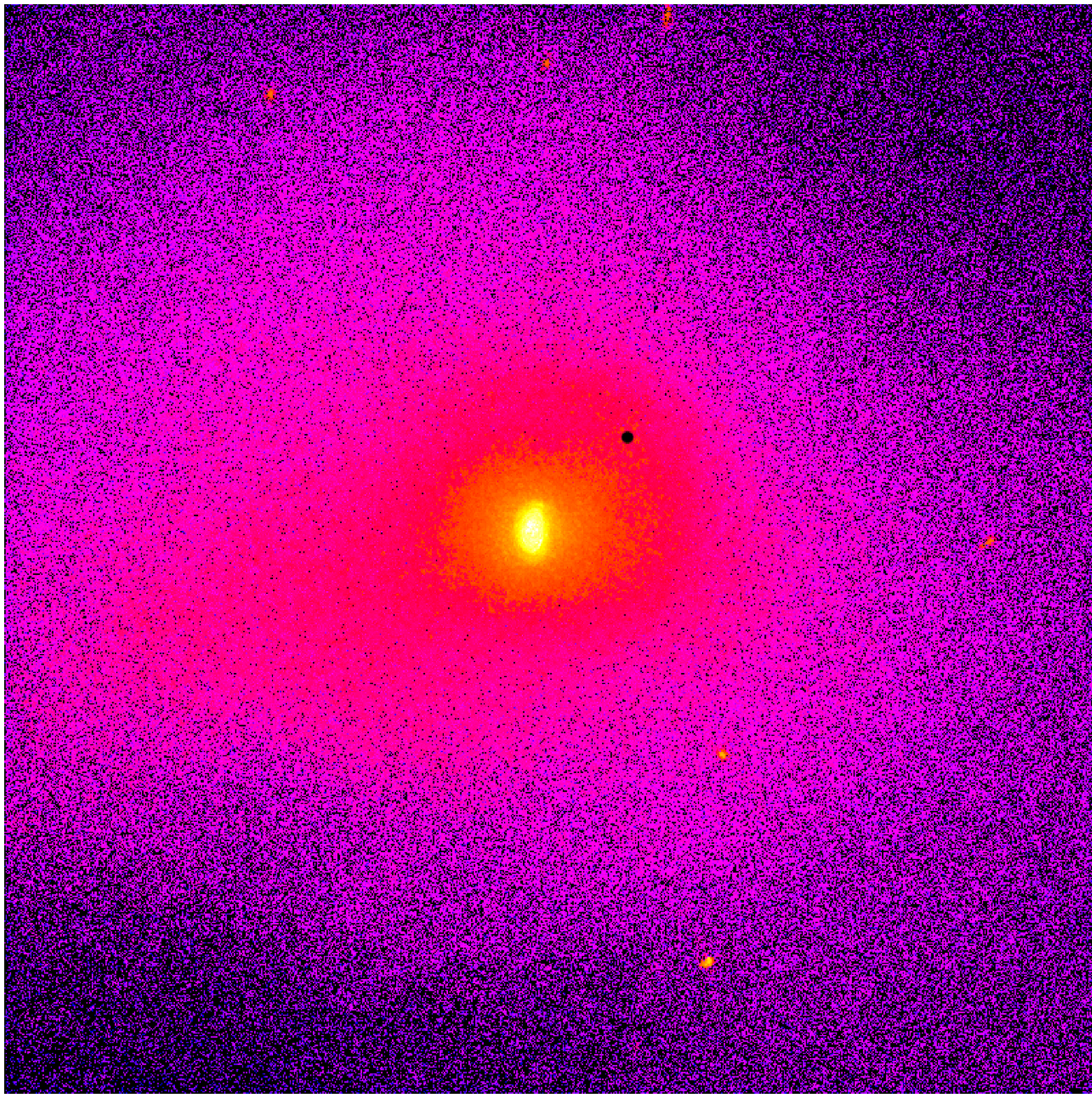}
\includegraphics[width=0.45\columnwidth,angle=0]{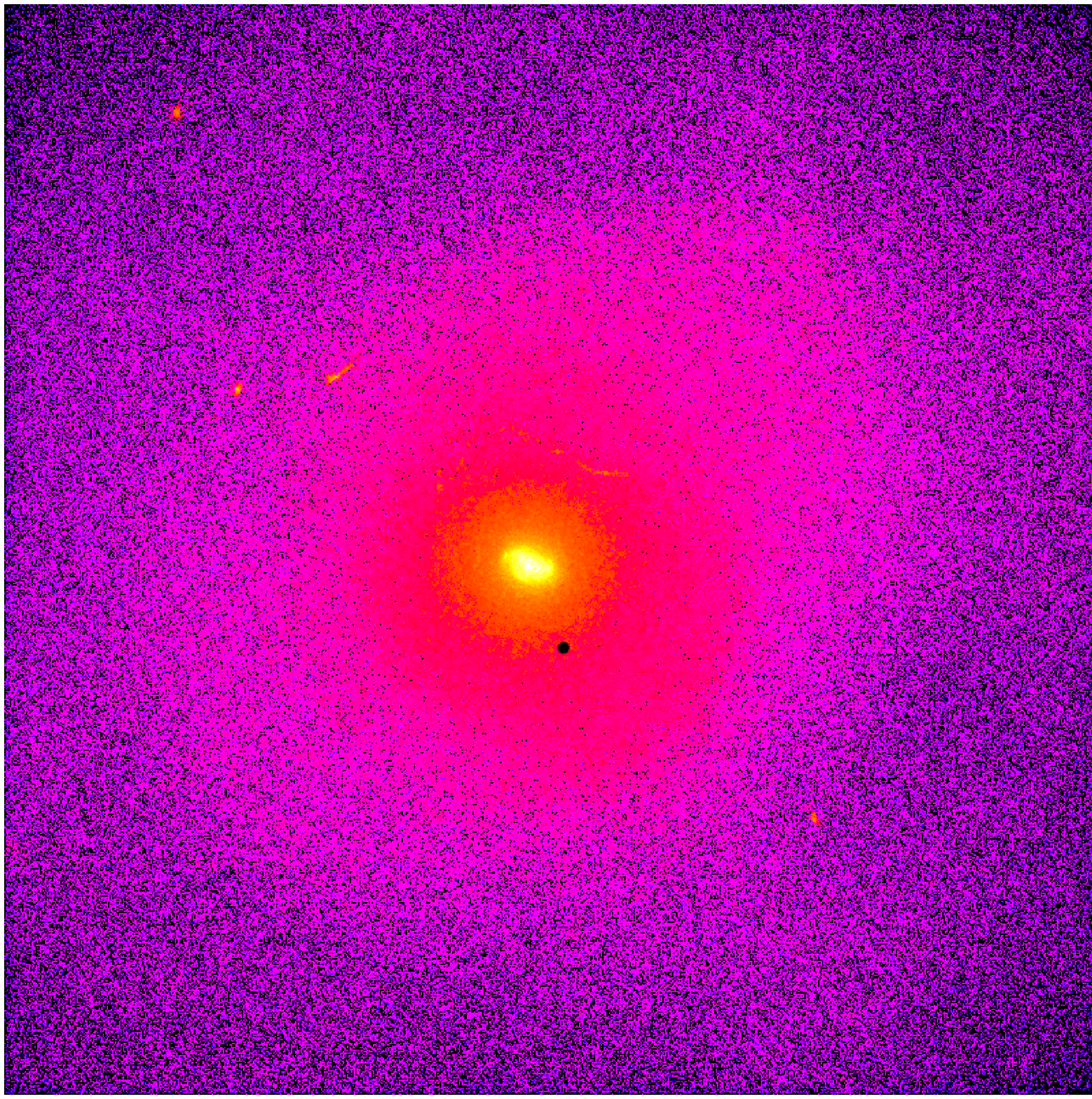}
\vskip 1.0mm
\begin{overpic}[width=0.90\columnwidth,angle=0]{ncp2013_colourbar.eps}
\put (1,3) {\textcolor{white}{$10^1$}}
\put (23,3) {$10^3$}
\put (48,3) {$10^5$}
\put (72,3) {$10^7$}
\put (95,3) {$10^9$}
\end{overpic}
\vspace{20pt}
\caption[Stellar density snapshots]{Time sequence (stellar density snapshots -- top to bottom, left to right -- in increments of 10~Myr) of the nuclear coup in the 1:4 coplanar, prograde--prograde merger, around $t = 1.45$~Gyr. The scale of the first snapshot is 8~kpc, that of the second and third is 4~kpc, and that of the last three is 2~kpc. A black dot marks BH$_1$, that is left `naked' after the disruption of $N_1$ in the sixth snapshot. The colour bar shows the (logarithmic) density scale in units of $2.2 \times 10^5$~M$_{\odot}$~kpc$^{-3}$.}
\label{ncp2013:fig:nuclear_coup_sequence}
\end{figure}

During a close encounter between the merging nuclei, rapidly varying gravitational fields inject energy into the systems. These gravitational shocks can lower the central density by redistributing mass to larger radii or completely unbinding material \citep{Ostriker1972,Spitzer1987,Gnedin1999,Taylor2001}. Unlike tidal stripping, which operates on the orbital time-scale of the material being stripped, tidal heating can inject energy even during very fast encounters.

During a fast encounter between a perturbing system of mass $M_{\rm p}$ and a shocked system of mass $M_{\rm s}$ with relative velocity $V$, the total energy injected into the shocked system is given by \citep{binney1987}

\begin{equation}\label{ncp2013:eq:delta_E_s}
\Delta E_{\rm s} = \frac{4G^2M_{\rm p}^2M_{\rm s}}{3V^2b^4}U(b/r_{\rm h})<r^2>,
\end{equation}

\noindent where $b$ is the impact parameter of the encounter and $<r^2>$ is the mass-weighted mean square radius of particles in the shocked system. $U(b/r_{\rm h})$ is a function that accounts for encounters where the two systems interpenetrate and the perturber cannot be approximated by a point mass. $r_{\rm h}$ represents the half-mass radius of the perturbing system. When the impact parameter is small compared to the half-mass radius, the total energy injected is reduced. We use the values of $U(b/r_{\rm h})$ given in \cite{binney1987}, approximating the density profiles of the systems as spherical \citet{Hernquist1990} profiles.

We compare the energy injected through tidal heating to the binding energy of the nuclei. We estimate the binding energy, $E_{\rm bind}$, as the energy required to move all the material in the nucleus to the edge of the nucleus, $r_{\rm nuc}$. This does not represent the energy required to completely unbind the nuclear material from the potential well of the merged galaxy. It instead approximates the energy required to smooth out the most highly bound portions of the nucleus. $E_{\rm bind}$ is given by

\begin{equation}\label{ncp2013:eq:E_bind}
E_{\rm bind} = \int_0^{r_{\rm nuc}}{4 \pi r'^2 \rho(r') [\phi(r_{\rm nuc}) - \phi(r')] dr'},
\end{equation}

\noindent where $\phi(r)$ is the gravitational potential of the shocked system at radius $r$. A dense nucleus has a large binding energy that is resistant to tidal heating. Additionally, a dense, centrally concentrated nucleus has a large mass as a perturber and small half-mass radius $r_{\rm h}$, increasing the energy injected into the other galaxy's nucleus.

\renewcommand{\arraystretch}{1.3}
\begin{table*}
\centering
\caption[]{Peak SFRs and results of the mergers. SFRs are the peak rates between the first pericentre passage and the merger of the nuclei ($N_1$ and $N_2$). Peak rates for each galaxy ($G_1$ and $G_2$) are SFRs within the central 100~pc. The binary time-scale is estimated using equation \eqref{ncp2013:eq:tau_DF} \citep[from][]{Colpi99} from the time of disruption of $N_1$ and/or $N_2$. The number in parenthesis is the approximate time from the beginning of the galaxy merger to the time of disruption.}
\begin{tabular}{l l l l l l l}
\hline
Simulation                   &  Global SFR & $G_1$'s SFR & $G_2$'s SFR & $N_1$ & $N_2$ & Binary time-scale \\
                                      &  $({\rm M}_{\odot}$~yr$^{-1})$ & $({\rm M}_{\odot}$~yr$^{-1})$ & $({\rm M}_{\odot}$~yr$^{-1})$ & Survival & Survival & (Myr) \\ [0.5ex]
\hline
Coplanar, prograde--prograde    &                       &                         &                               &                & & \\
1:2                                                                     & 18.9               & 1.77               & 3.63                      & No          & Yes & 13.2 (+1300) \\
1:4                                                                     & 8.3                 & 0.19               & 7.65                       & No          & Yes & 23 (+1500) \\
1:6                                                                     & 4.35               & 0.06               & 3.78                      & No          & Yes & 17.4 (+2000) \\
1:10                                                                   & 1.1                 & 0.12               & 0.73                      & Yes         & No & $>$92 (+3000) \\\\
Inclined                         &                        &                        &                              &                 & & \\
1:2                                                                      & 9.44               & 1.82              & 8.9                        & No           & Yes & 18.3 (+1300) \\
1:4                                                                      & 1.96               & 0.28              & 0.32                      & Yes         & No & 660 (+1700) \\\\
Coplanar, retrograde &                        &                       &                               &                 & & \\
1:2 (retrograde--prograde)                                                         & 11.4               & 3.49              & 2.34                      & No           & No & $<$8.3 (+1300)\\
1:2 (prograde--retrograde)                                                          & 26.9               & 4.8                 & 0.93                      & Yes         & No & 223 (+1200)\\ [1ex]
\hline
\end{tabular}
\vspace{10pt}
\label{ncp2013:tab:1}
\end{table*}
\renewcommand{\arraystretch}{1.0}

Due to the strong dependence of the tidal heating on the impact parameter, $b$, the initial pericentre passages inject little energy into the nuclei compared to the total binding energy. The energy becomes important when the nuclei pass within $r \leq 100$~pc with typical velocities of $V \simeq 300$--500~km~s$^{-1}$. During these encounters, the energy injected from the companion nucleus can be greater than $E_{\rm bind}$, causing the nucleus to be disrupted and leaving the central BH `naked' \citep[see also][]{Governato1994}, without any bound gas or stars.

Following the third pericentre passage in the 1:4 coplanar, prograde--prograde merger, $N_2$ is significantly denser than $N_1$ (Fig.~\ref{ncp2013:fig:central_density_1to4_cop_pro_pro}). During the fourth and fifth pericentre passages, when the nuclei pass within $\simeq$100~pc of each other, tidal shocks reduce $G_1$'s central density. At the sixth pericentre passage, the nuclei pass within $\leq$29~pc of each other with a relative velocity of 415~km~s$^{-1}$ and $N_1$ is unbound. The relatively less dense $N_1$ injects far less energy into $N_2$, which survives the encounter intact, and remains at the centre of the merger remnant where the last and strongest burst of star formation of the merger occurs. The primary BH (BH$_1$), now without any bound stars or gas, is left on an elliptical orbit around the merger remnant with an apocentre of 230~pc.

The occurrence of the nuclear coup can be effectively shown in Fig.~\ref{ncp2013:fig:BH_potential_vs_time}, in which we plot the gravitational potential of the two BHs as a function of time, from right before the third pericentre passage onwards. Around fourth pericentre, the gravitational potential of BH$_1$ becomes higher than that of BH$_2$, clearly indicating that BH$_2$ is now in a deeper potential well (the remnant centre) and BH$_1$ is now orbiting it. The nuclear coup can also be visualized via a time sequence of stellar density snapshots (Fig.~\ref{ncp2013:fig:nuclear_coup_sequence}), around the same time shown in Fig.~\ref{ncp2013:fig:BH_potential_vs_time}, in increments of 10~Myr. In the sixth snapshot, BH$_1$ is clearly `naked', after the disruption of $N_1$.


\section{Impact of mass ratio and orbital parameters}\label{ncp2013:sec:Results}

In this section, we assess the impact of different mass ratios and orbital parameters. The results are summarized in Table \ref{ncp2013:tab:1} and discussed further in Section~\ref{ncp2013:sec:Discussion}, specifically in light of BH pairing and binary formation.


\subsection{Impact of mass ratio}\label{ncp2013:sec:Impact_of_mass_ratio}

\begin{figure}
\centering
\vspace{5pt}
\includegraphics[width=1.0\columnwidth,angle=0]{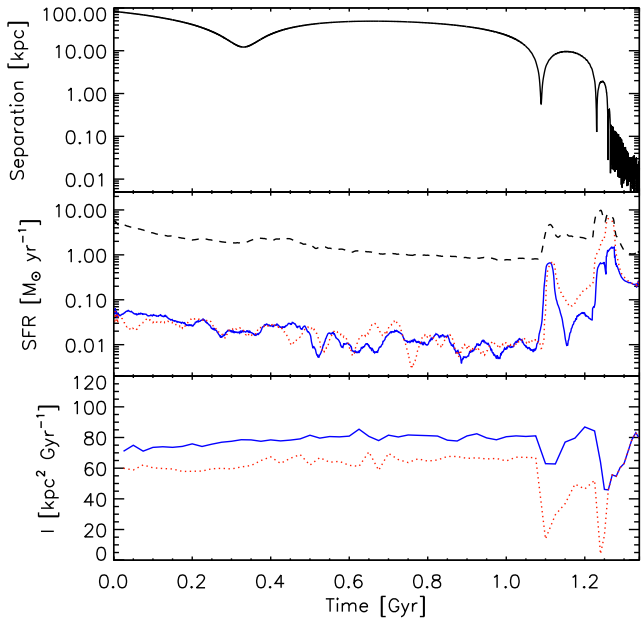}
\vspace{6pt}
\caption[Results of the 1:2 coplanar, prograde--prograde simulation]{Results of the 1:2 coplanar, prograde--prograde simulation. Top panel: separation between the central BHs of each galaxy. Middle panel: global SFR across both galaxies (black, dashed line), and central ($<$100~pc) SFR of $G_1$ (blue, solid line) and $G_2$ (red, dotted line). Bottom panel: angular momentum per unit mass of gas in the central kpc of $G_1$ (blue, solid line) and $G_2$ (red, dotted line). All quantities are shown as a function of time.}
\label{ncp2013:fig:3panels_1to2_cop_pro_pro}
\end{figure}

\begin{figure}
\centering
\vspace{5pt}
\includegraphics[width=1.0\columnwidth,angle=0]{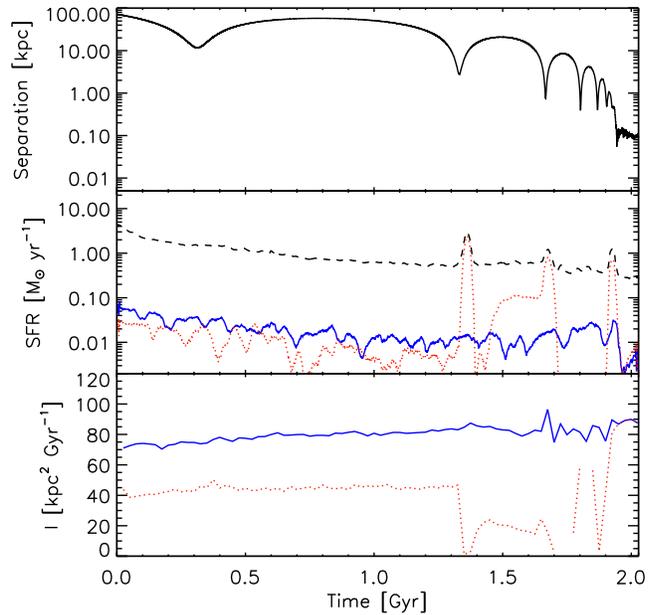}
\vspace{6pt}
\caption[Results of the 1:6 coplanar, prograde--prograde simulation]{Results of the 1:6 coplanar, prograde--prograde simulation. Top panel: separation between the central BHs of each galaxy. Middle panel: global SFR across both galaxies (black, dashed line), and central ($<$100~pc) SFR of $G_1$ (blue, solid line) and $G_2$ (red, dotted line). Bottom panel: angular momentum per unit mass of gas in the central kpc of $G_1$ (blue, solid line) and $G_2$ (red, dotted line). All quantities are shown as a function of time.}
\label{ncp2013:fig:3panels_1to6_cop_pro_pro}
\end{figure}

\begin{figure}
\centering
\vspace{5pt}
\includegraphics[width=1.0\columnwidth,angle=0]{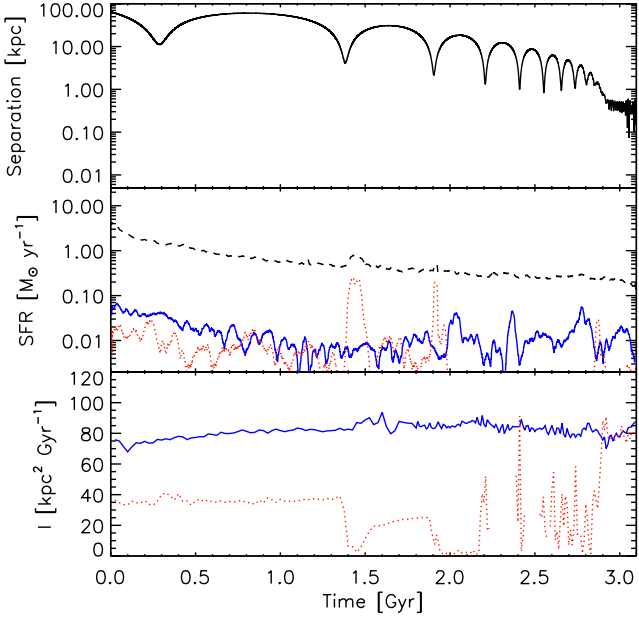}
\vspace{6pt}
\caption[Results of the 1:10 coplanar, prograde--prograde simulation]{Results of the 1:10 coplanar, prograde--prograde simulation. Top panel: separation between the central BHs of each galaxy. Middle panel: global SFR across both galaxies (black, dashed line), and central ($<$100~pc) SFR of $G_1$ (blue, solid line) and $G_2$ (red, dotted line). Bottom panel: angular momentum per unit mass of gas in the central kpc of $G_1$ (blue, solid line) and $G_2$ (red, dotted line). All quantities are shown as a function of time.}
\label{ncp2013:fig:3panels_1to10_cop_pro_pro}
\end{figure}

In this section, we compare the results of the coplanar, prograde--prograde mergers (mass ratios 1:2, 1:4, 1:6, and 1:10). Figures \ref{ncp2013:fig:3panels_1to4_cop_pro_pro}, \ref{ncp2013:fig:3panels_1to2_cop_pro_pro}, \ref{ncp2013:fig:3panels_1to6_cop_pro_pro}, and \ref{ncp2013:fig:3panels_1to10_cop_pro_pro} show the evolution of these mergers. We find that $N_2$ is able to form a dense central cusp and disrupt $N_1$ in the 1:2, 1:4, and 1:6 coplanar, prograde--prograde mergers, but not in our 1:10 merger. The formation of a dense cusp depends on the strength of gaseous inflows and the ability of $G_2$'s gas to survive direct collisions with $G_1$'s disc. The strongest nuclear star formation occurs in $G_2$ in our 1:4 run, then becomes weaker as the mass ratio decreases and $G_2$ loses more gas to ram pressure stripping from $G_1$. In the following, we discuss the detailed findings, first by galaxy and then by merger.

\begin{figure}
\centering
\vspace{5pt}
\includegraphics[width=1.0\columnwidth,angle=0]{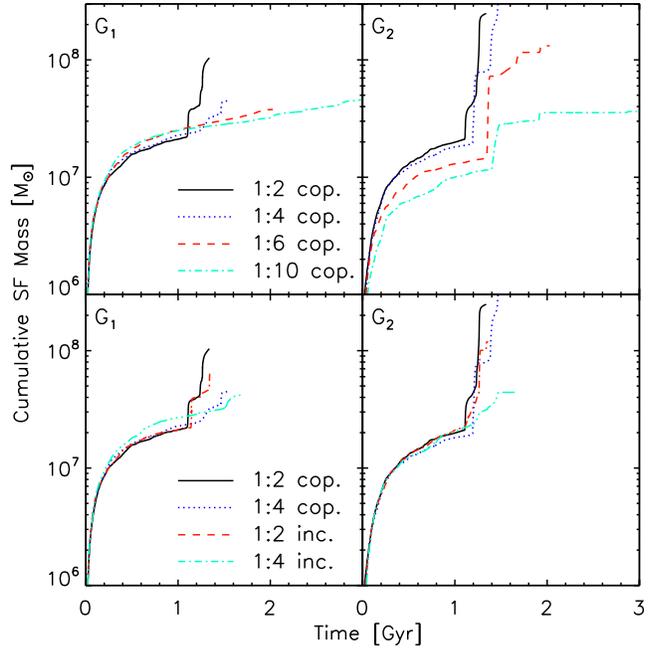}
\vspace{6pt}
\caption[]{Cumulative mass in new star formation in the central 100~pc of $G_1$ (left-hand panels) and $G_2$ (right-hand panels). Top panels: all coplanar, prograde--prograde mergers (1:2: black, solid line; 1:4: blue, dotted line; 1:6: red, dashed line; and 1:10: cyan, dot--dashed line). Bottom panels: coplanar, prograde--prograde mergers (1:2: black, solid line; 1:4: blue, dotted line) and inclined mergers (1:2: red, dashed line; 1:4: cyan, dot--dashed line).}
\label{ncp2013:fig:cum_mass_in_new_SF}
\end{figure}

\begin{enumerate}

\item {\it Primary galaxy ($G_1$)}. As the mass ratio of the merger decreases, $G_1$ experiences weaker tidal torques due to the relatively less massive $G_2$. The result is a more limited loss of angular momentum, down to no loss at all in the smallest mass ratios, and a lack of strong merger-induced star formation. The top-left panel of Fig.~\ref{ncp2013:fig:cum_mass_in_new_SF} shows the cumulative mass in new stars formed in the central 100~pc of $G_1$ in each coplanar, prograde--prograde merger. $G_1$ shows a strong central burst of star formation at second and third pericentre in the 1:2 merger and a weaker enhancement following second pericentre in the 1:4 merger, driven by a weak bar, but no response in the 1:6 and 1:10 runs. The peak nuclear SFR prior to the merger of the nuclei is shown in Table \ref{ncp2013:tab:1} for each run. The global peak star formation decreases with mass ratio, as does the peak response of $G_1$ down to a minimum peak rate of 0.1--0.2~M$_{\odot}$~yr$^{-1}$.

\item {\it Secondary galaxy ($G_2$)}. The tidal response of $G_2$, on the other hand, grows stronger as the mass ratio decreases and $G_1$ becomes relatively more massive. This leads to stronger inflows, but strong nuclear star formation depends on dense central gas surviving the collision between the gaseous discs. Figs \ref{ncp2013:fig:3panels_1to6_cop_pro_pro} and \ref{ncp2013:fig:3panels_1to10_cop_pro_pro} show that the strongest loss of angular momentum at second pericentre occurs in the 1:6 and 1:10 mergers. However, as the gas mass and density of $G_2$'s disc decrease, the disc is more strongly affected by ram pressure from $G_1$'s disc, and the mass of the dense star-forming clump generally decreases with mass ratio. The exception are the 1:2 and 1:4 runs, where the total mass in central gas that survives the disc interaction is similar. In the 1:4 run, however, the gas is more strongly compressed during the disc collision. The gas therefore reaches higher densities and fuels a stronger burst of star formation.

\end{enumerate}

\begin{enumerate}

\item {\it 1:2 merger.} As a result of the strong burst of star formation in both $G_1$ and $G_2$ at second pericentre in the 1:2 merger (Fig.~\ref{ncp2013:fig:3panels_1to2_cop_pro_pro}), the nuclei have similar central densities. Stronger angular momentum loss and inflows in $G_2$ at third pericentre fuel a large increase in its central mass. As in the 1:4 merger, $N_1$ is completely disrupted due to tidal heating from $N_2$ during the fourth and fifth pericentre passages.

\item {\it 1:6 merger.} In the 1:6 merger, $G_2$'s disc is strongly affected by ram pressure from $G_1$'s disc. This limits the amount of cold, dense gas available for star formation. At third pericentre, ram pressure removes the majority of the gas. Supernova feedback then expels the remaining gas, leaving $G_2$ completely gas poor. The remaining evolution is slower than in the 1:2 and 1:4 mergers, resulting in more pericentre passages before the nuclei merge. Tidal heating reduces the central mass and density of $N_2$ during these passages while $N_1$ remains intact. Despite the effects of tidal heating, $N_2$ remains significantly denser than $N_1$. Eventually, $N_2$'s orbit circularizes within the disc of $G_1$, then plunges inward towards $N_1$, which is disrupted during the plunge when the nuclei pass within $\leq$55~pc of each other. BH$_1$ is left on a circular orbit around the merger remnant with a radius of $\simeq$100~pc.

\item {\it 1:10 merger.} The 1:10 merger proceeds similarly to the 1:6 merger. $G_2$ loses its gas to ram pressure following the third pericentre passage and experiences the weakest star formation of the prograde--prograde mergers. When the orbit of $G_2$ circularizes within $G_1$'s disc, $G_2$ is only denser than $G_1$ on scales of $\simeq$15--20~pc. Despite the lack of significant merger-induced star formation in $G_1$, $G_2$ is unable to build up enough central mass to survive. During the plunge (passing within $\simeq$400~pc of the centre of $G_1$), $N_2$ is disrupted down to its dense central cusp which has a total mass of $10^7$~M$_{\odot}$, an order of magnitude more than the mass of BH$_2$. After this time the cusp remains on an elliptical orbit with an apocentre of $\simeq$550~pc for $\sim$200~Myr. The pericentres get closer and closer with time, until they reach $\simeq$70~pc. We stop the simulation at this point (after more than 3~Gyr total running time). If the orbit of the cusp is able to decay further, we estimate it is not dense enough to survive a direct encounter with $N_1$. Using equations \eqref{ncp2013:eq:delta_E_s} and \eqref{ncp2013:eq:E_bind}, we estimate that $N_2$'s cusp would be completely disrupted upon passing within $\simeq$30~pc of the centre of $N_1$.

\end{enumerate}

\subsection{Impact of orbital parameters}\label{ncp2013:sec:Impact_of_orbital_parameters}

We supplement the study of the coplanar, prograde--prograde mergers with inclined and coplanar, retrograde mergers. In brief, in the inclined mergers tidal torques are weaker, and it is more difficult for $G_2$ to build a strong nuclear cusp. A nuclear coup occurs in the 1:2 case, but in the 1:4 case $N_2$ is instead disrupted. In the 1:2 coplanar, retrograde mergers ram pressure is stronger, and the compression triggers high nuclear SFRs in $G_1$, making it more difficult for $G_2$ to build a central cusp denser than that of $G_1$. In the case where $G_1$'s spin axis is flipped, both $N_2$ and $N_1$ are disrupted, when instead $G_2$'s spin axis is flipped, $N_2$ is disrupted by tidal shocks.


\subsubsection{Inclined orbits}

We summarize here the results of our inclined mergers (mass ratios 1:2 and 1:4), in which the disc of $G_1$ is tilted $\ang{45}$ with respect to the orbital plane. $G_2$'s disc is unchanged compared to the coplanar, prograde--prograde mergers. In the inclined mergers, $G_2$ feels weaker tidal torques from $G_1$ during the second pericentre passage than in coplanar mergers, resulting in only a weak enhancement in the central SFR. Instead of a burst, we see sustained star formation at an order of magnitude higher SFR than during the early quiescent phase of the merger. This enhancement is fed by low angular momentum gas, previously stripped from both galaxies during the second pericentre passage, which now reforms the disc of $G_2$. The main increase in central mass in $G_2$ occurs during third pericentre, when the reformed disc is compressed by the ram pressure of the $G_1$'s disc. As the mass ratio of the merger decreases, the reformed disc is less massive and is strongly stripped during the third pericentre passage, preventing $G_2$ from efficiently forming stars and building a dense central cusp.

\begin{figure}
\centering
\vspace{5pt}
\includegraphics[width=1.0\columnwidth,angle=0]{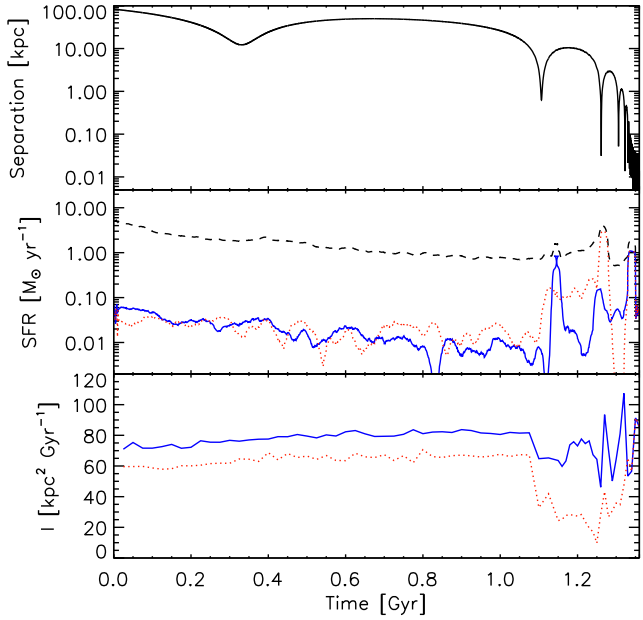}
\vspace{6pt}
\caption[Results of the 1:2 inclined simulation]{Results of the 1:2 inclined simulation. Top panel: separation between the central BHs of each galaxy. Middle panel: global SFR across both galaxies (black, dashed line), and central ($<$100~pc) SFR of $G_1$ (blue, solid line) and $G_2$ (red, dotted line). Bottom panel: angular momentum per unit mass of gas in the central kpc of $G_1$ (blue, solid line) and $G_2$ (red, dotted line). All quantities are shown as a function of time.}
\label{ncp2013:fig:3panels_1to2_inclined}
\end{figure}

\begin{figure}
\centering
\vspace{5pt}
\includegraphics[width=1.0\columnwidth,angle=0]{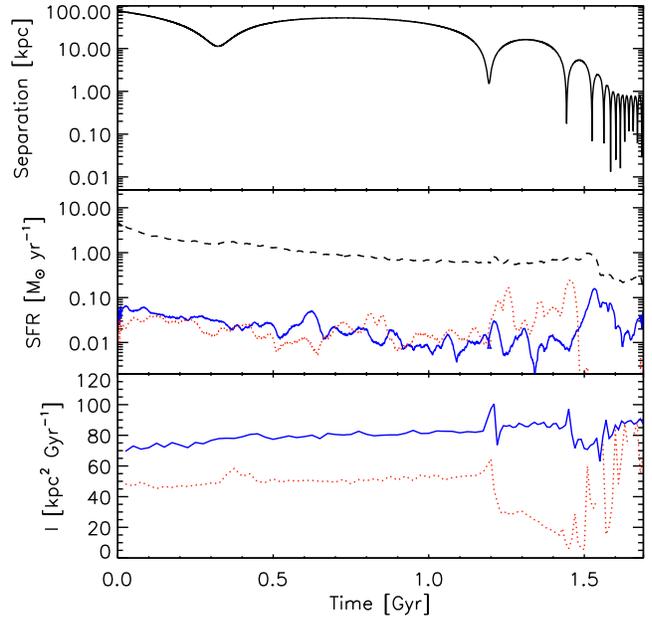}
\vspace{6pt}
\caption[Results of the 1:4 inclined simulation]{Results of the 1:4 inclined simulation. Top panel: separation between the central BHs of each galaxy. Middle panel: global SFR across both galaxies (black, dashed line), and central ($<$100~pc) SFR of $G_1$ (blue, solid line) and $G_2$ (red, dotted line). Bottom panel: angular momentum per unit mass of gas in the central kpc of $G_1$ (blue, solid line) and $G_2$ (red, dotted line). All quantities are shown as a function of time.}
\label{ncp2013:fig:3panels_1to4_inclined}
\end{figure}

\begin{enumerate}

\item {\it 1:2 inclined merger.} Fig.~\ref{ncp2013:fig:3panels_1to2_inclined} shows the evolution of the 1:2 inclined merger. The results of the simulation are very similar to the results of the coplanar, prograde--prograde run. At second pericentre, the angle between the discs of the galaxies produces a weaker tidal torque on $G_2$'s disc than in the coplanar case, leading to a smaller reduction in angular momentum and no significant burst of star formation. Instead, $G_2$ experiences sustained nuclear star formation at a rate of 0.1--0.2~M$_{\odot}$~yr$^{-1}$ until third pericentre, fed by low angular momentum gas falling back into the nucleus after being stripped during the interaction with $G_1$'s disc. The resulting reformed disc is less massive than in the coplanar case, but it is smaller and denser. As a result, it is strongly compressed during the third pericentre passage and hosts a nuclear starburst reaching 8.9~M$_{\odot}$~yr$^{-1}$, higher than in the coplanar merger. During the fifth and sixth pericentre passages, $N_1$ is disrupted by $N_2$.

\item {\it 1:4 inclined merger.} As in the 1:2 inclined merger, weak angular momentum loss and gaseous inflows lead to little enhancement in star formation at second pericentre in the 1:4 inclined merger (Fig.~\ref{ncp2013:fig:3panels_1to4_inclined}). $G_2$'s gas disc, strongly stripped during the encounter with $G_1$'s disc, reforms with predominantly low angular momentum material, leading to a slow reduction in the average angular momentum of gas in the central kpc (bottom panel of Fig.~\ref{ncp2013:fig:3panels_1to4_inclined}). This low angular momentum gas fuels nuclear star formation, but the SFR remains low and it does not contribute significantly to the formation of a dense central cusp. The reformed disc is significantly less massive and dense than the reformed disc in the 1:2 inclined merger. As a result, much of the disc is stripped due to ram pressure during the third apocentre. The remaining gas is compressed and efficiently forms stars, but the SFR remains low and there is again no significant increase in central density in $G_2$. Supernova feedback removes the rest of the gas following third apocentre. During subsequent pericentre passages, the central density of $G_2$ decreases due to energy injection from tidal shocks. At sixth pericentre, $N_2$ is disrupted by $N_1$, which survives the encounter. BH$_2$ orbits the merger remnant on an elliptical orbit with an apocentre of 750~pc.

\end{enumerate}

The bottom panels of Fig.~\ref{ncp2013:fig:cum_mass_in_new_SF} show a comparison in the cumulative nuclear star formation between the coplanar and inclined mergers. The total star formation in the 1:2 mergers is very similar despite the change in inclination, although the triggering of the star formation is different, as discussed above. The inclination plays a much larger role in influencing the star formation in $G_2$ in the 1:4 mergers (solid versus dashed green lines, bottom-right panel), where there is almost an order of magnitude difference in the cumulative star formation between the mergers. This shows why $G_2$ is unable to develop the dense central cusp necessary to disrupt $N_1$ in the 1:4 inclined case.


\subsubsection{Retrograde orbits}

We also consider coplanar mergers that are retrograde, where the spin axes of the two galaxies have the opposite direction. Both of the mergers we consider here have a mass ratio of 1:2. The coplanar, retrograde--prograde run is similar in setup to the 1:2 coplanar, prograde--prograde merger, except the spin axis of $G_1$ has been flipped with respect to the orbital angular momentum vector of the galaxies. In the coplanar, prograde--retrograde merger, the spin axis of $G_2$ has been flipped. The retrograde orbits lead to stronger ram pressure in the disc interaction compared to the prograde--prograde mergers because the impacting retrograde galaxy adds the rotational velocity to its orbital velocity vector. The stronger interaction produces high nuclear SFRs in $G_1$, making it more difficult for $G_2$ to build a central cusp denser than that of $G_1$. In the coplanar, retrograde--prograde run, $N_2$ sustains enough star formation to become similarly dense to $N_1$, causing both nuclei to be disrupted late in the merger. The formation of a massive bridge in the coplanar, prograde--retrograde merger prevents $G_2$ from reforming a significant gaseous disc after second pericentre. $N_2$ therefore remains less dense than $N_1$ and is disrupted by tidal shocks.

\begin{figure}
\centering
\vspace{5pt}
\includegraphics[width=1.0\columnwidth,angle=0]{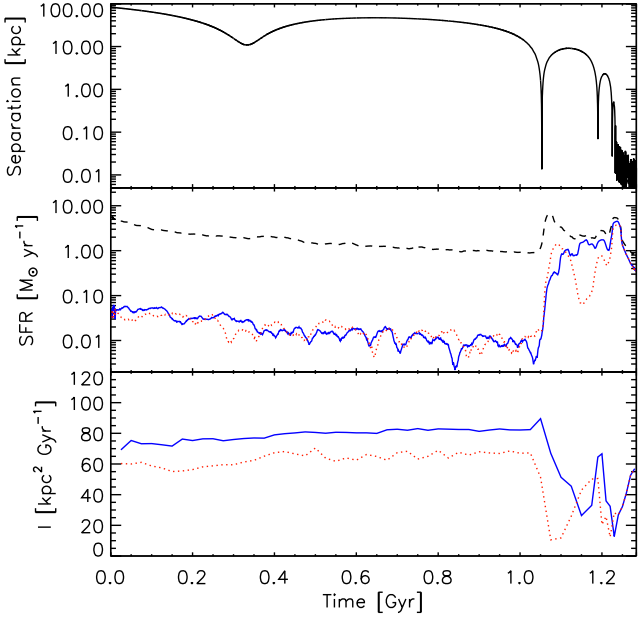}
\vspace{6pt}
\caption[Results of the 1:2 coplanar, retrograde--prograde simulation]{Results of the 1:2 coplanar, retrograde--prograde simulation. Top panel: separation between the central BHs of each galaxy. Middle panel: global SFR across both galaxies (black, dashed line), and central ($<$100~pc) SFR of $G_1$ (blue, solid line) and $G_2$ (red, dotted line). Bottom panel: angular momentum per unit mass of gas in the central kpc of $G_1$ (blue, solid line) and $G_2$ (red, dotted line). All quantities are shown as a function of time.}
\label{ncp2013:fig:3panels_1to2_cop_ret_pro}
\end{figure}

\begin{figure}
\centering
\vspace{5pt}
\includegraphics[width=1.0\columnwidth,angle=0]{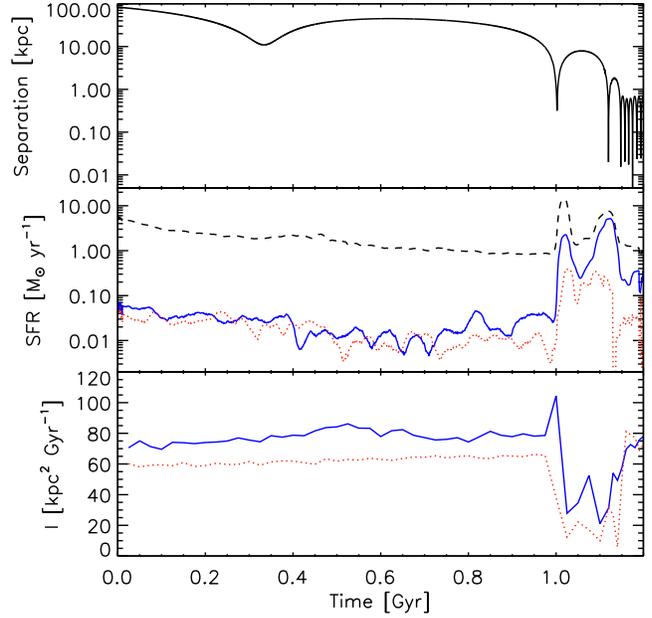}
\vspace{6pt}
\caption[Results of the 1:2 coplanar, prograde--retrograde simulation]{Results of the 1:2 coplanar, prograde--retrograde simulation. Top panel: separation between the central BHs of each galaxy. Middle panel: global SFR across both galaxies (black, dashed line), and central ($<$100~pc) SFR of $G_1$ (blue, solid line) and $G_2$ (red, dotted line). Bottom panel: angular momentum per unit mass of gas in the central kpc of $G_1$ (blue, solid line) and $G_2$ (red, dotted line). All quantities are shown as a function of time.}
\label{ncp2013:fig:3panels_1to2_cop_pro_ret}
\end{figure}

\begin{enumerate}

\item {\it Coplanar, retrograde--prograde merger.} $G_2$ in the coplanar, retrograde--prograde merger evolves similarly to $G_2$ in the 1:2 coplanar, prograde--prograde merger. Strong inflows at second pericentre and compression during the disc interaction produce a high central gas density, leading to a nuclear starburst (Fig.~\ref{ncp2013:fig:3panels_1to2_cop_ret_pro}). After the second pericentre passage, supernova feedback heats the gas, preventing further strong star formation as $G_2$'s gaseous disc reforms. At third pericentre, the gas is again compressed, producing another starburst that increases the central mass and density of $N_2$. $G_1$ does not experience a strong starburst following second pericentre, but forms a strong bar following the encounter, which is not present in the prograde--prograde merger. The bar funnels gas into the centre of $G_1$, leading to a higher sustained nuclear SFR than in $G_2$ near apocentre. The result of this nuclear star formation is that both nuclei are similarly dense when they merge. During the fifth pericentre passage, when the nuclei pass within 11~pc of each other, tidal heating unbinds both nuclei. The central BHs of both galaxies are left orbiting around the merger remnant, which is largely made up of new stars that formed in the final starburst.

\item {\it Coplanar, prograde--retrograde merger.} The coplanar, prograde--retrograde interaction between the discs in this merger leads to a strong shock in the disc gas. The leading edge of each galaxy is rotating into the disc collision, increasing the relative velocity of the impact. This shocked gas forms a massive bridge between the galaxies as they approach apocentre following the second pericentre passage. In the coplanar, retrograde--prograde merger, this shocked gas passes around the nucleus of $G_1$. In the coplanar, prograde--retrograde merger, however, the shocked gas passes directly through $G_1$'s nucleus, strongly compressing the central gas there. This interaction leads to the strongest global SFR in any of the mergers presented here, peaking at 26.9~M$_{\odot}$~yr$^{-1}$. Much of this star formation occurs in the massive bridge that links the galaxies, but the central star formation in $G_1$ is higher than in $G_2$ (Fig.~\ref{ncp2013:fig:3panels_1to2_cop_pro_ret}). The strong star formation in the gaseous bridge and following supernova feedback prevents the gas from reforming $G_2$'s disc. $G_2$'s disc therefore remains low in mass and hosts little star formation during the third pericentre passage. $G_1$'s nucleus sustains a consistently higher SFR than $N_2$ and remains denser. At fourth pericentre, $N_2$ is disrupted during a close encounter with $N_1$. BH$_2$ is left on an elliptical orbit with an apocentre of $\simeq$650~pc.

\end{enumerate}


\section{Discussion}\label{ncp2013:sec:Discussion}


\subsection{Black hole pairing and binary evolution}\label{ncp2013:sec:BH_pairing_and_binary_evolution}

Our simulations allow us to follow the dynamics of BHs in merging galaxies from scales of hundreds of kpc to $\sim$10~pc. We can accurately track the motion of the nuclei of the merging galaxies and study how tidal effects and merger-induced star formation affect the formation of the BH pair in a realistic environment, and for a variety of mass ratios and orbital configurations. Technically, we do not resolve separations where the BHs become bound and form a binary. Using the definitions described in Section \ref{ncp2013:sec:Introduction}, $a_{\sigma}=7$--8~pc and $a_{\rm M}=14$--16~pc for all our mergers, where we have used the BH and stellar bulge quantities from the initial conditions. Using instead the final masses of the BHs at the end of our simulations, the stated values increase by at most 50 per cent. Our resolution is therefore not sufficient to track the orbits of the BH down to the binary stage. However, it is sufficient to reliably track the orbits down to the formation of a close pair and we then extrapolate analytically their further evolution, as explained below.

We find that gas dynamics and star formation are very important to the successful formation of a BH pair in minor mergers, via the formation of a nucleus that `delivers' the BHs to the central region of the remnant \citep{Yu2002, stelios2005, Callegari2009}. \citet{MerrittCruz2001} also studied stellar-only minor mergers between giant elliptical galaxies and dwarf galaxies with relatively steeper central density profiles. If BHs are excluded from the galaxies, they find that the secondary's cusp survives the merger intact, significantly increasing the central density of the merged galaxy. If BHs are included, however, tidal heating from BH$_1$ reduces the central density of $N_2$ on small scales and the central density of the merged galaxy is only increased slightly. BH$_1$ is less massive in our mergers and is less important dynamically in $N_1$. It contributes negligibly to the tidal shock on $N_2$ during the final pericentre passages in our runs where $N_1$ is disrupted.

The disruption of a nucleus delays the formation of a binary BH system, since the dynamical friction time-scale for a `naked' BH is longer than that for the original nucleus.

Following disruption, we estimate the time-scale for the orbit of the BH to decay and reach the centre of the merged galaxy analytically. This is because the effects of dynamical friction in our simulations may be underestimated on a lone BH due to gravitational softening on small scales. In our galaxies, the separation at which a binary forms is $<$30~pc, therefore our resolution is not sufficient to track the orbits down to the binary stage. To estimate the time for a BH binary to form, we consider the effects of dynamical friction acting on the `naked' BH as it moves through the merger remnant as proposed by Colpi et al. (1999\nocite{Colpi99}), who study the decay in the orbits of satellites in $N$-body simulations: 

\begin{equation}\label{ncp2013:eq:tau_DF}
\tau_{\rm DF} = 1.2\frac{J_{\rm cir}r_{\rm cir}}{\left(GM_{\rm sat}/e\right)\ln{\left(M_{\rm halo}/M_{\rm sat}\right)}} \epsilon^{0.4}.
\end{equation}

Here $M_{\rm sat}$ and $M_{\rm halo}$ are the masses of the satellite and halo, respectively. $J_{\rm cir}$ and $r_{\rm cir}$ are the orbital angular momentum and radius of a circular orbit with the same energy as the initial orbit of the satellite. $\epsilon$ is the ratio of the angular momentum of the initial orbit to $J_{\rm cir}$. This parameter accounts for the faster decay of elliptical orbits, which pass deeper into the halo and encounter higher background densities, increasing the force of dynamical friction. $e$ accounts for mass loss from the satellite due to tidal stripping as the orbit decays. In the case of rigid satellites such as BHs, $e=1$.

In determining $\tau_{\rm DF}$, we first calculate the energy per unit mass of the orbit of the BH using $E/M = (1/2) v^2 + \Phi$, where $v$ is the velocity of the BH relative to the centre of mass and $\Phi$ is the gravitational potential per unit mass of the BH. We then move outward from the centre of mass of the merger remnant until we find a circular orbit with the same energy. The angular momentum and radius of this orbit determine $J_{\rm cir}$, $r_{\rm cir}$, and $\epsilon$. We set $M_{\rm halo}$ equal to the total mass enclosed within this circular orbit, and $M_{\rm sat}$ equal to the mass of the BH.

Since equation \eqref{ncp2013:eq:tau_DF} depends inversely on the mass of the satellite body, the loss of the material surrounding BH$_1$ increases the dynamical friction time-scale, delaying the formation of a binary compared to a case where both BHs retain their cusps throughout the decay. However, the same scaling implies also that a nuclear coup paints a more optimistic case than when $N_2$ is disrupted, as in that case the dynamical friction time-scale depends on the {\it smaller} mass of BH$_2$. In the 1:10 merger we provide a lower limit to the binary formation time-scale assuming that the remainders of $N_2$ are disrupted at 30~pc from the centre. However, we do not directly witness this event and therefore simply assume this is a strict lower limit. 

In our simulations where $G_2$ is unable to sustain strong central star formation, the `naked' BH$_2$ is left at a separation of $>$500~pc, significantly delaying the formation of a BH binary (see binary formation time-scales in Table \ref{ncp2013:tab:1}). Without any surrounding stars and gas, BH$_2$ sinks more slowly due to dynamical friction. Additionally, the BH spends most of its orbit far from the centre of the merger remnant where the ambient density is low and dynamical friction is inefficient. When $G_2$ does build a dense cusp throughout the merger, $N_2$ survives the merger down to the centre of $G_1$. When $N_1$ is disrupted (nuclear coup), the `naked' BH$_1$ is left orbiting very close to the remnant. Dynamical friction is more effective than in the previous case because BH$_1$ is more massive than BH$_2$ and because the BH is left orbiting in a denser environment. The BHs quickly reach the resolution limit of the simulation, near separations where they will form a binary. Still, it is important to consider the interaction between the nuclei when estimating the overall time-scale for BHs to coalesce. Even when both nuclei survive down to small scales in the merger, the following formation of a BH binary is not instantaneous. 

In summary, we find that the pairing time-scale increases with the mass ratio of the merger, as expected (numbers in parenthesis in Table \ref{ncp2013:tab:1}), and broadly speaking the mass ratio is also the main parameter that determines binary formation. However, at fixed mass ratio, the details of binary formation depend on nuclear dynamics, on scales $<$50~pc, which in turn are determined by effective nuclear star formation. For instance, we find that for most of our 1:2 mergers a binary forms on relatively short time-scales, except for the prograde--retrograde case. Taking into account pairing and binary time-scales, the orbital decay from hundreds of kpc to pc scale takes in all cases less than a Hubble time: the time from $z=3$ to 0 is 11.5~Gyr.


\subsection{Nuclear star formation and disruption}\label{ncp2013:sec:nuclear_SF}

Although $G_1$ is the more massive galaxy in our mergers, our results show that its nucleus $N_1$ can be disrupted in a variety of mass ratios if the discs are coplanar and both prograde. As a result, the central baryonic material of the remnant comes mainly from $N_2$. However, this orientation maximizes the tidal response of the disc and the strength of the following starburst \citep{MihosHernquist1996,Cox2008}. Indeed, our inclined mergers produce weaker starbursts. As the mass of the two galaxies becomes more equal, the inclination seems to play less of a role in determining the strength of inflows in the disc interaction. Our 1:2 coplanar, prograde--prograde merger and 1:2 inclined merger produce similar results (Fig.~\ref{ncp2013:fig:cum_mass_in_new_SF}), with $N_2$ disrupting $N_1$ in both runs. Tilting $G_1$'s disc in a more minor merger makes a large difference; the star formation in $G_2$ in the 1:4 inclined merger is far weaker than in the coplanar, prograde--prograde merger and $N_1$ is no longer disrupted. Our exploration of the possible orbital parameters is by no means exhaustive, but we have shown that $N_2$ can grow to be as dense as $N_1$ for several disc orientations in a major merger (1:2).

An important aspect of each merger is the collision between the gaseous discs. Ram pressure during the second pericentre passage removes much of the gas in $G_2$, leaving a massive gaseous bridge linking the galaxies. The survival of dense nuclear gas through the second pericentre and the formation of a new disc at apocentre are vital to producing a further starburst at third pericentre, at which point the nuclei have completed the majority of their star formation. We find that the gaseous disc that reforms in $G_2$ following second pericentre flips in angular momentum compared to the original in all our coplanar, prograde--prograde mergers except for the 1:10 case. Unfortunately, it is difficult to analytically follow the interaction between the discs and determine the cause of the spin flip. The spin direction depends on the angular momentum of the gaseous bridges and tidal arms that feed $G_2$'s disc. It is also difficult to determine how the spin flip affects star formation during the third pericentre passage, when $G_2$'s disc again collides with $G_1$ and its disc takes on the spin direction of the more massive $G_1$. The strongest burst of star formation in any of our mergers occurs in the 1:2 coplanar, prograde--retrograde merger, suggesting that a prograde--retrograde encounter may be the most violent and lead to a strong starburst. This effect may enhance the SFR at third pericentre in our 1:2, 1:4, and 1:6 coplanar, prograde--prograde mergers, where the spin flip following second pericentre has made $G_2$'s disc now retrograde. We plan to study the influence and implications of the spin flip further in future work.

While we have focused on merger-driven starbursts in our simulations, the majority of the star formation occurs during the early quiescent phase before the gas discs collide. As \cite{Cox2008} found, the starbursts do not efficiently convert a large amount of the global gas into stars in unequal-mass mergers. While the global conversion of gas into stars is dominated by the initial phase, the starbursts contribute preferentially to the central region, where quiescent star formation contributes negligibly to the mass build-up.

Observations of paired galaxies in the Sloan Digital Sky Survey agree with our result that in unequal-mass mergers the smaller galaxy experiences stronger star formation. \cite{Woods2007} consider 3613 galaxies in pairs and split them into minor and major pairs based on their relative magnitude. The major pairs (with a difference in $z$ magnitude $\Delta m_{z} < 2$) show signs of ongoing star formation in both galaxies. The minor pairs show signs of active star formation only in the less massive galaxy. Additionally, the activity in the galaxies increases at small separations. Accordingly, we find that both galaxies in our 1:2 mergers exhibit strong central star formation, whereas in our 1:6 and 1:10 mergers, only $G_2$ experiences significant merger-induced star formation. The interacting system of NGC 7770 and NGC 7771 (stellar mass ratio 1:10) also shows an enhancement in star formation only in the less massive galaxy \citep{AH2012}.


\subsection{Influence of resolution}\label{ncp2013:sec:Influence_of_resolution}

In order to follow the build-up of central mass and the following dynamical interaction of the nuclei, numerical simulations must resolve very small scales. In the 1:4 coplanar, prograde--prograde merger, $G_2$ experiences much stronger nuclear star formation than $G_1$, but is denser only on scales of $\leq$75~pc when the nuclei begin to interact. Without high resolution on scales of tens of pc, the density contrast between the nuclei could not be studied. Additionally, tidal heating becomes strong enough to disrupt the nuclei only when they pass each other on scales $\leq$50~pc, scales that are unresolved in most studies of interacting galaxies. For example, \citet{Robertson2005}, in studying the evolution of BH scaling relations during mergers, used a redshift-dependent gravitational softening for baryonic particles, equal to $\sim$140~pc at $z=0$. In a similar study, Johansson et al. (2009\nocite{Johansson2009}) used a gravitational softening of $\sim$30~pc for baryonic and BH particles and of $\sim$120~pc for dark matter particles. \citet{Younger2008}, when simulating the self-regulated growth of BHs through major and minor mergers, and disc instabilities, have a spatial resolution of 30--50~pc. \citet{Cox2008}, in their study on the effect of mass ratio on merger-driven starbursts, did not include BHs, and used a gravitational softening of 100 and 400~pc for baryonic and dark matter particles, respectively. We note that these papers were not focused on the dynamical evolution, and therefore did not, effectively, need the same level of detail that we required.

The minimum gas temperature is 10~K in the simulations presented here. During starbursts, many of the new stars form out of gas at temperatures of 10--100~K. However, at these low temperatures and at the densities at which typical stars form, the gas structure is not resolved. The smoothing length of the gas becomes smaller than the softening length at low temperatures, inhibiting further collapse \citep[see discussion in][]{Bate97}, but the Jeans mass contains only a few particles. To test the impact of gas cooling on our results, we ran an additional 1:4 coplanar, prograde--prograde merger with a gas temperature floor of 500~K, where the gas remains well resolved. The overall evolution of the merger is similar, although we see somewhat stronger central star formation in both galaxies than with a lower temperature floor as inflowing cold gas penetrates further into the galaxy before forming stars. The outcome of the merger is unchanged in this simulation and, in particular, the nuclear coup occurs as in the simulation with a lower temperature floor.


\section{Conclusions}\label{ncp2013:sec:Conclusions}

We present simulations of unequal-mass galaxy mergers, where $G_1$ is the larger galaxy, and $G_2$ is the smaller, focusing on the spatial distribution of merger-triggered starbursts and the consequences for the dynamics of the central nuclei ($N_1$ and $N_2$) and BHs. We will discuss accretion and the triggering of active galactic nuclei in a forthcoming paper. We summarize our findings below.

\begin{enumerate}

\item We find that $G_2$ generally experiences stronger nuclear star formation than $G_1$. In some mass ratios and orientations, its nucleus, $N_2$, becomes denser on small scales and disrupts $N_1$. The disruption is consistent with tidal heating due to fast collisions between the nuclei at separations of $\leq$50~pc.

\item The survival of $N_2$ during the merger depends on the interaction between the gaseous discs of the galaxies. If $G_2$ has a high central gas mass and deep potential well to resist ram pressure, the gas will be compressed during the collision with $G_1$'s disc, driving strong star formation. The majority of the nuclear star formation occurs following second and third pericentre. In order to sustain significant star formation during third pericentre, $G_2$ must recapture gas that was stripped by $G_1$. 

\item As the mass ratio of the merger decreases, $G_2$'s disc is less massive and is more strongly affected by ram pressure from $G_1$'s disc. Ram pressure therefore removes much of the gas in $G_2$, limiting the amount of central gas that is able to form stars.

\item If $G_2$ is able to form a dense central cusp, it is more resistant to heating from tidal shocks and retains a larger bound central mass, sinking further due to dynamical friction and leading more quickly to the formation of a close BH pair on scales of 10--100~pc. When $N_1$ is disrupted, we analytically find that the binary formation time-scale is fast, occurring in less than 100~Myr. In mergers where $N_2$ is instead disrupted due to insufficient central star formation, the formation of a binary is delayed (Table \ref{ncp2013:tab:1}). We conclude that it is vital to follow star formation and the interaction between the nuclei on scales less than 100~pc in order to accurately understand the formation and evolution of BH binaries.

\end{enumerate}

\section*{Acknowledgements}
MV acknowledges funding support from NASA, through award ATP NNX10AC84G, from SAO, through award TM1-12007X, from NSF, through award AST 1107675, and from a Marie Curie Career Integration grant (PCIG10-GA-2011-303609). This work was granted access to the HPC resources of TGCC under the allocation 2013-t2013046955 made by GENCI. This research was supported in part by the National Science Foundation under grant no. NSF PHY11-25915, through the Kavli Institute for Theoretical Physics and its program `A Universe of Black Holes'. PRC thanks the Institut d'Astrophysique de Paris for hosting him during his visits.

\scalefont{0.94}
\setlength{\bibhang}{1.6em} 
\setlength\labelwidth{0.0em}
\bibliographystyle{mn2e}
\bibliography{nuclearcoups_20140127}
\normalsize

\end{document}